\documentclass[a4paper]{jpconf}
\usepackage{graphicx}
\usepackage{subeqnarray}
\usepackage{mathptmx}
\usepackage{amsfonts}
\usepackage[centertags]{amsmath}
\usepackage{amssymb}

\def\ba{\begin{array}}
\def\ea{\end{array}}
\def\be{\begin{equation}}
\def\ee{\end{equation}}
\def\bea{\begin{eqnarray}}
\def\eea{\end{eqnarray}}
\usepackage{wasysym}
\usepackage{MnSymbol}
\usepackage[hypertex]{hyperref}
\usepackage[verbose,colorlinks=true,naturalnames=true,linkcolor=blue]{}

\newcounter{rown}

\def\sec{\setcounter{equation}0}
 
\begin{document}

\title{Real and pseudoreal forms of  D=4 complex Euclidean \newline
(super)algebras  
 and   super--Poincar\'{e}/ 
  super--Euclidean  r-matrices }

\author{A.~Borowiec$^{1)}$,   J.~Lukierski$^{1)}$\footnote{Presented by J.~Lukierski}  and  V.N.~Tolstoy$^{2)}$}
\address{
$^{1)}$Institute for Theoretical Physics, 
University of Wroc{\l }aw, pl. Maxa Borna 9,  
50--205 Wroc{\l }aw, Poland
\\
$^{2)}$Lomonosov Moscow State University, 
Skobeltsyn Institute of Nuclear Physics, 
Moscow 119991, Russian Federation}



\begin{abstract}
We provide the classification of real forms of complex D=4 Euclidean algebra 
 $\mathcal{\epsilon}(4; \mathbb{C})=\mathfrak{o}(4;\mathbb{C}))\ltimes\mathbf{T}_{\mathbb{C}}^4$
 as well as (pseudo)real forms of  complex D=4 Euclidean superalgebras
$\mathcal{\epsilon}(4|N; \mathbb{C})$ for N=1,2.
Further we present our  results: N=1 and  N=2 supersymmetric D=4 Poincar\'{e}  and Euclidean r-matrices obtained  by using 
 D= 4 Poincar\'{e}  r-matrices provided by Zakrzewski \cite{Z94}  
 For N=2 we shall consider the general superalgebras with two central charges.
\end{abstract}

\section{Introduction}

The classification of quantum deformations of Lorentz symmetries described by classical r-matrices was given firstly by Zakrzewski \cite{Z97} (see also \cite{Mudrov}), and has been further extended to the classification of  classical r-matrices for Poincar\'{e} algebra in \cite{Z94}. The classification of dual Hopf-algebraic quantum deformations of Poincar\'{e} group were presented in \cite{PW}. Subsequently, the infinitesimal r-matrix description of the deformations of Poincar\'{e} algebra presented in \cite{Z94} has been extended in several papers to finite Hopf-algebraic deformations, with conclusive results obtained in \cite{T07}. Because majority of studied deformations were triangular, i.e. described by the twist deformations, they permitted (see e.g. \cite{Bloch, Kulish}) to derive explicit formulas for the non-commutative  algebra of deformed space-time coordinates  by the use of so-called star product realizations.

The study of deformations of spacetime supersymmetries and the corresponding deformed superspaces were less systematic, related mostly either with the supersymmetrization of  simplest Abelian canonical twist deformation of Poincar\'{e} symmetries
\cite{KS}--\cite{IS}
or with the supersymmetric extension of $\kappa$-deformation \cite{LNS,KLMS}. The supersymmetrization of  such a canonical twist $F$ for $N=1$ Poincare superalgebra looks as follows
\begin{equation}\label{1a1}
 F=\exp{{1\over 2}\theta^{\mu\nu}P_\mu\wedge P_\nu} \quad \rightarrow
 \quad
  {\mathcal F}= \exp{{1\over 2}\theta^{\mu\nu}P_\mu\wedge
  P_\nu}\,
  \exp \xi^{\alpha \beta} Q_\alpha\wedge Q_\beta \, ,
 \end{equation}
 where $\theta_{\mu\nu} = - \theta_{\nu\mu}$, $\xi^{\alpha \beta}=\xi^{\beta\alpha}$ and 
 $\hat{x}\wedge \hat{y}\equiv (\hat{x}\otimes \hat{y})- (-1)^{|x||y|} (\hat{y} \otimes\hat{x})$ where $|x|= 0,1$ is the $Z_2$-grading of the superalgebra element $\hat{x}$ with definite grading.  
The classification  of the supersymmetric triangular deformations for finite-dimensional simple Lie superalgebras was firstly given in a mathematical framework by one of the present authors \cite{T3}; one should add also that already long time ago the nontriangular Drinfeld-Jimbo  deformation has been provided for all complex finite-dimensional simple Lie superalgebras in \cite{T4,CKNT,KT6}  but without discussion of its real forms.   Recently however 
(see \cite{T5,BLMT})  the classification of Poincar\'{e} r-matrices in \cite{Z97} has been extended to $D=4,N=1$ Poincare and  Euclidean \cite{BLMT} supersymmetries, which provided  physically   important deformations of non-semisimple Lie superalgebras.
 New result in this paper is the presentation of the list of D=4 N=2 classical super-Poincar\'{e} and super-Euclidean r-matrices.

 We add that  there were also considered twisted Euclidean $N=1$ and $N=2$ supersymmetries (see e.g.  \cite{Zup}) but only under the assumption that the fermionic part of the twist factor was a supersymmetric enlargement of canonical Abelian twist (see (\ref{1a1})). 
 In our recent paper by considering partial classification of $N=1$ complex $r-$matrices and their pseudoreal (Euclidean) and real (Poincar\'{e}) forms
  \cite{BLMT} we have obtained large class of $N=1$ supersymmetric twists which provide new  D=4 superspaces with Lie-algebraic deformation of bosonic spacetime sector. 

\indent
 In quantum deformations  approach  the basic primary notion is the Hopf-algebraic deformation of (super)symmetries which subsequently implies the modification of (super)spacetime algebra.
It follows  that for classifying the possible deformations of (super)symmetries one should list the deformations of corresponding Hopf (super)algebras.
In non-supersymmetric case the most complete discussion of such Hopf-algebraic approach to field theories deformed by canonical twists $F$ (see (\ref{1a1}))   was presented in
\cite{FW}; for Poincar\'{e} and Euclidean supersymmetries such way of introducing deformations via supertwist, firstly advocated in \cite{KS,Zup}, was presented recently in \cite{BLMT,Seib}.  In present paper we  supplement  classical
  $N=2$   super--Poincar\'{e} $r-$matrices which generate the corresponding N=2 supersymmetric twist (triangular) deformations. Subsequently using $*-$product formulation
	  \cite{Bloch,Kulish} one can  provide effective formulae for the description of quantum-deformed N=2 
superspaces. We add that the particular supersymmetric $N=2$ twist deformations considered  in earlier studies \cite{FerraraMacia}--\cite{Buch} 
 appear as particular cases in the list of  N=2 deformations presented in this paper.

In Sect.~2 and~3 we shall consider  in some detail inhomogeneous
  $\mathfrak{o}(4;\mathbb{C})$  algebra with its real forms and the N=1,2 superextensions;  in Sect.~4 and~5 we present our partial results describing N=1,2 supersymmetric Poincar\'{e} and Euclidean  r-matrices.
More detailed plan of our paper is the following:  
    In Sect. 2 we shall consider complex $\mathfrak{o}(4)$  and inhomogeneous 
		$\mathfrak{io}(4)$  algebras and its $N=1$ SUSY extension. We shall also provide their real and pseudoreal forms defining  $D=4$ Poincare,  Euclidean and Kleinian algebras and corresponding $N=1$ real/pseudoreal superalgebras. 
In Sect. 3 we shall discuss complex $N=2$   superalgebras with two central charges and odd sector described by $8$ independent complex supercharges ($Q_\alpha^{\ \ a}, Q_{\dot{\alpha}\, b}$,\ $\alpha, \dot{\alpha} = 1,2 \ a,b=1,2$).
 Further following \cite{YuManin}--\cite{BerVNT} and by using suitable conjugations or pseudoconjugations 
 (real or pseudoreal forms) we shall describe  $N=2$ real Poincar\'{e} and Kleinian superalgebras as well as 
   complex
 selfconjugate Euclidean superalgebras.
  In order to compare with earlier results for N=0 (nonsupersymmetric case) and N=1 we shall
 provide 
 in Sect.~4  
 the tables of classical Poincar\'{e} (super)-r-matrices given in 
 \cite{Z94} and \cite{BLMT}\footnote{We point out here that among 21   classical $r-$matrices obtained by Zakrzewski~\cite{Z94} there is one class which satisfies modified YB equation. We shall consider 20 classes of $r-$matrices which lead to triangular deformations.}.  
 	   In Sect.~5 we present new results: we  use  the set of real  Poincar\'{e}
  classical r-matrices provided 
by Zakrzewski in \cite{Z94}, and  describe the ones which do have N=2 supersymmetric extension;
	we present also N=2 \ D=4 Euclidean supersymmetric r-matrices.
	 Finally  in Sect.~6 we present final remarks. 

In Appendix A  we shall outline the general theory of conjugations and pseudoconjugations for Lie superalgebras.

 The list of  $N=2$  complex supersymmetric  r-matrices   and  their Kleinian 
$\mathfrak{o}(2,2)$ real counterparts satisfying suitable (pseudo)reality conditions  will be presented in our next publications.

\section{Complex D=4  Euclidean algebra, its simple (${N\negthickspace=\negthickspace1}$)   supersymmetrization and their real or pseudoreal forms}

\subsection{Complex  $\mathfrak{io}(4,\mathbb{C})$ algebra and its real forms}
\sec
It is known that there are four  real forms  of complex orthogonal Lie algebras  $\mathfrak{o}(4,\mathbb{C})$  corresponding to three different nondegenerate $D=4$ metric signatures and the fourth one  which can be
obtained by imposing quaternionic  structure (see e.g.  \cite{Gilmore-a,BarutRaczka}):

i) $\mathfrak{o}(4)\cong \mathfrak{o}(3)\oplus \mathfrak{o}(3) \cong \mathfrak{su}(2)\oplus \mathfrak{su}(2)$  \qquad \qquad -- Euclidean case

ii) $\mathfrak{o}(3, 1)
\cong
\mathfrak{sl}(2, \mathbb{C})_\mathbb{R}$   \  \  \  \qquad \quad\qquad \qquad\qquad   -- Lorentzian case\footnote{The real generators $\mathfrak{o}(3,1)$ can be described by complex--conjugated  generators  of  $\mathfrak{o}(3;\mathbb{C})$ and  $\overline{\mathfrak{o}(3;\mathbb{C})}$ (see (\ref{bch1})).}

iii) $\mathfrak{o}(2,2)\cong \mathfrak{o}(2,1)\oplus \mathfrak{o}(2,1) \cong \mathfrak{sl}(2, \mathbb{R})\oplus
\mathfrak{sl}(2, \mathbb{R})  \cong$

$\mathfrak{su}(1,1)\oplus \mathfrak{su}(1,1) \cong \mathfrak{sp}(2, \mathbb{R})\oplus \mathfrak{sp}(2, \mathbb{R})$
$\qquad$  \  \ -- Kleinian case

iv)  $\mathfrak{o}^*(4)\cong \mathfrak{o}(3)\oplus \mathfrak{o}(2,1) \cong \mathfrak{su}(2)\oplus \mathfrak{su}(1,1)$
$\quad$\, -- quaternionic case.\\
The quaternionic origin of the fourth real form follows from isomorphism with orthogonal quaternionic algebras \cite{Gilmore-a}
\begin{equation}\label{quater1}
\mathfrak{o}(2;\mathbb{H})=\mathfrak{o}^*(4) ,   \qquad \mathfrak{o}(1;\mathbb{H})=\mathfrak{o}(2) \, .
 \end{equation}
In first three cases (Euclidean, Lorentzian, Kleinian) one can lift the corresponding  real forms  to inhomogeneous algebra
$\mathcal{\epsilon}(4; \mathbb{C})=\mathfrak{o}(4;\mathbb{C}))\ltimes\mathbf{T}_{\mathbb{C}}^4$, where the complex Abelian
generators  describe complexified momentum fourvectors ($\mathcal{P}_\mu\in \mathbf{T}_{\mathbb{C}}^{4}$).

In fourth case one can introduce the momenta as complex $\mathfrak{o}^*(4)$ vectors described by
 second order fourcomponent  $SU(2)\times SU(1,1)$  complex spinors. The corresponding inhomogeneous algebra
$\mathfrak{o}^*(4)\ltimes\mathbf{T}_{q}^4$  can be endowed with quaternionic structure if we perform
the following contraction of the quaternionic symmetric coset (see also (\ref{quater1}))
 \begin{equation}\label{quater2}
\mathfrak{o}(3;\mathbb{H})=\mathfrak{o}(2;\mathbb{H})\oplus \mathfrak{o}(1;\mathbb{H})\oplus
\frac{\mathfrak{o}(3;\mathbb{H})}{\mathfrak{o}(2;\mathbb{H})\oplus \mathfrak{o}(1;\mathbb{H})}\Rightarrow
(\mathfrak{o}^*(4)\oplus\mathfrak{o}(2))\ltimes\mathbf{T}_{q}^4 \, ,
\end{equation}
where $\mathbf{T}_{q}^4$  has eight real dimensions, i.e. the dimensionality of $\mathbf{T}_{\mathbb{C}}^4$
is not reduced. We add that the inhomogeneous $\mathfrak{o}^*(4)$ algebra does not play   known  significant role in the
description of physically relevant $D=4$   geometries.

The most known real form of $\mathfrak{io}(4,\mathbb{C})$ is the $D=4$ Poincar\'{e} Lie algebra ${{p}}(3,1))=\mathfrak{o}(3,1)\ltimes
\mathbf{T}^{1,3}$ generated by the
Poincar\'{e} fourmomenta $P_\mu\in \mathbf{T}^{1,3}$
and  six Lorentz rotations
$L_{\mu\lambda}\in\mathfrak{o}(3,1)$ 
 ($\mu,\nu=1,2,3,4$) and looks as follows:

\begin{eqnarray}\label{2a1}
\begin{array}{rcccl}
[L_{\mu\nu},\,L_{\lambda\rho}]\!\! & =\!\! & i\bigl(g_{\nu\lambda}\,L_{\mu\rho}-
g_{\nu\rho}\,L_{\mu\lambda}+g_{\mu\rho}\,L_{\nu\lambda}-g_{\mu\lambda}\,
L_{\nu\rho}\bigr)~,\qquad L_{\mu\nu}\!\! & =\!\! & -L_{\nu\mu}~,
\\[12pt]
[L_{\mu\nu},\,P_{\rho}]\!\! & =\!\! & i\bigl(g_{\nu\rho}\,P_{\mu}-
g_{\mu\rho}\,P_{\nu}\bigr)~,\qquad\qquad\qquad\qquad\quad\;
[P_{\mu},\,P_{\nu}]\!\!&=\!\!&0~,
\end{array}%
\end{eqnarray}
where $g_{\mu\nu}=\mathop{\rm diag}\,(-1,-1,-1,1)$ is the Minkowski (Lorentzian) metric.
If we replace in (\ref{2a1})  such a  metric by the Euclidean one i.e. $g^E_{\mu\nu}=-\delta_{\mu\nu}=
\mathop{\rm diag}\,(-1,-1,-1,-1)$, one gets the $D=4 $ Euclidean algebra
${\mathcal{\epsilon}}(4)=\mathfrak{o}(4) \ltimes\mathbf{T}^4$, described by Euclidean generators $\mathcal{P}_\mu,\ \mathcal{L}_{\mu\nu}$:
\begin{eqnarray}\label{2a3}
\begin{array}{rcccl}
[\cal{L}_{\mu\nu},\,\cal{L}_{\lambda\rho}]\!\! & =\!\! & -i\bigl(\delta_{\nu\lambda}\,\cal{L}_{\mu\rho}-
\delta_{\nu\rho}\,\cal{L}_{\mu\lambda}+\delta_{\mu\rho}\,\cal{L}_{\nu\lambda}-\delta_{\mu\lambda}\,
\cal{L}_{\nu\rho}\bigr)~,\qquad \cal{L}_{\mu\nu}\!\! & =\!\! & -\cal{L}_{\nu\mu}~,
 \\[12pt]
[\cal{L}_{\mu\nu},\,\cal{P}_{\rho}]\!\! & =\!\! & -i\bigl(\delta_{\nu\rho}\,\cal{P}_{\mu}-
\delta_{\mu\rho}\,\cal{P}_{\nu}\bigr)~,\qquad\qquad\qquad\qquad\quad\;
[\cal{P}_{\mu},\,\cal{P}_{\nu}]\!\!&=\!\!&0~.
\end{array}%
\end{eqnarray}
The Poincar\'{e} algebra can be obtained from Euclidean one by the following substitution ($r,s,t=1,2,3$)
\begin{equation}\label{2a2}
   L_{\mu\nu}=(\mathcal{L}_{rs},\   
	i \mathcal{L}_{0r})\quad   , \qquad   P_\mu=(\mathcal{P}_r , \ \ 
	i \mathcal{P}_0) 
\end{equation}
or equivalently $M_r=\mathcal{M}_r$, $N_r=- 
i \mathcal{N}_r$, where $M_r={1\over 2}\epsilon_{rst}L_{st}, \ \mathcal{M}_r={1\over 2}\epsilon_{rst}\mathcal{L}_{st}$ and $N_r=L_{4r}$,  $\mathcal{N}_r=\mathcal{L}_{4r}$.
 The imaginary unit occurring in (\ref{2a2})  effectively change in (\ref{2a1}) 
 the Euclidean  metric  $g^E_{\mu\nu}$ into the  Lorentzian one $g_{\mu\nu}$.

If we choose the metric with Kleinian (neutral)
  signature $g^K_{\mu\nu}=\mathop{\rm diag}\,(1,-1,1,-1)$ 
  one gets Lie algebra $\mathfrak{io}(2,2))=\mathfrak{o}(2,2)\ltimes
\mathbf{T}^{2,2}$, which can be obtained by the following change of the Euclidean generators
\begin{equation}\label{2a2k}
M^K_i=  i \mathcal{M}_i,\quad N^K_i= i  \mathcal{N}_i\quad \mbox{(i=1,3)},
\quad M^K_2=   \mathcal{M}_2,\quad N^K_2=  \mathcal{N}_2 ,\quad
P^K_\mu =(\mathcal{P}_1, i \mathcal{P}_2, \mathcal{P}_3,  i\mathcal{P}_4) \, .
\end{equation}

For describing the $\mathfrak{o}^*(4)$ algebra with quaternionic structure one should split the real
$\mathfrak{o}(4)$ generators as follows
\begin{equation}\label{2a2q}
\mathfrak{o}(4)=\mathfrak{u}(2)\oplus \frac{\mathfrak{o}(4)}{\mathfrak{u}(2)}
\end{equation}
 and multiply the coset generators by "$i$".

In order to embed three cases  of inhomogeneous algebras into unified framework one can consider the complex
$D=4$ Euclidean algebra $\mathcal{\epsilon}(4; C)$ and introduce its three real forms (Euclidean, Poincar\'{e} and Kleinian). These real forms   are introduced with the help of three non-isomorphic  antilinear  involutive conjugations
 $I_A \to I^\#_A$  $(I_A\in ({\mathcal L}_{\mu\nu}, {\mathcal P}_{\mu}))$, where $\#=\dagger,\ \ddagger,\ \oplus$.
 The following reality conditions imposed on the generators of $\mathcal{E}(4;C)$ ($r,s=1,2,3 ; i=1,3; k=2,4$)
\begin{eqnarray}\label{2a4}
{\cal L}^\dagger_{r s} ={\cal L}_{r s},\quad {\cal L}_{4 r}^\dagger =-{\cal L}_{4 r},  \qquad {\cal P}_r^\dagger =  {\cal P}_r, \quad {\cal P}_4^\dagger = -{\cal P}_4 \qquad \hbox{Poincar\'{e} case}
\\[12pt]\label{2a5}
\cal{L}_{\mu\nu}^\ddagger =\cal{L}_{\mu\nu}, \qquad \cal{P}_{\mu}^\ddagger = \cal{P}_{\mu} \qquad \hbox{Euclidean case}
\\[12pt]\label{2a6}
{\cal L}^\oplus_{1\, 3} ={\cal L}_{1\, 3},\quad {\cal L}_{2\, 4}^\oplus ={\cal L}_{2\,4 },  \quad {\cal L}_{i\, k}^\oplus =-{\cal L}_{i\, k}\quad {\cal P}_i^\oplus =  {\cal P}_i, \quad {\cal P}_{k}^\oplus = -{\cal P}_{k} \qquad \hbox{Kleinian case}
\end{eqnarray}
define respectively real  $D=4$  Poincar\'{e}, Euclidean and Kleinian algebras.  We observe that first two real structures (conjugations) coincide on $\mathcal{E}(3)$ subalgebra.

Another convenient basis in complex $\mathcal{E}(4; \mathbb{C})$ algebra is obtained by introducing the pair of
chiral (left-handed) and anti-chiral (right-handed) generators: $2M^{\pm}_r=  {1\over 2 }\epsilon_{rst}{\mathcal L}_{st}\pm {\mathcal L}_{4\,r}\equiv {\mathcal M}_r\pm i {\mathcal N}_r$ describing  two complex commuting $\mathfrak{o}_\pm(3,C)\equiv sl_\pm(2,C)$ subalgebras
\begin{eqnarray}\label{bch1}
  [M^{\pm}_r, M^{\pm}_s] =  i \epsilon_{ijk} M^{\pm}_k,& 
  &[M^\pm_r, M^\mp_s]=0 \, ,
  \\[10pt]
   \label{bch1bis}
  \qquad [M^{\pm}_r, {\cal P}_{s}] = { i \over 2}(\epsilon_{rst}{\cal P}_t\mp \delta_{rs}{\cal P}_4),&   
 & [M^{\pm}_r, {\cal P}_{4}]=\pm{ i \over 2}{\cal P}_r \, .
\end{eqnarray}
The reality conditions  (\ref{2a4}--\ref{2a6}) 
  imposed on the complex   generators $M^\pm$ look as follows
 (we provide also the fourth reality condition related with quaternionic structure; $r=1,2,3; i=1,3$)
\begin{eqnarray}\label{bch2}
(M^{\pm}_r)^\dagger  = M^{\mp}_r   \qquad\qquad 
\qquad \hbox{Poincar\'{e} case}
\\[12pt]\label{bch3}
(M^{\pm}_r)^\ddagger =M^{\pm}_r   \qquad\qquad 
\qquad \hbox{Euclidean case}
\\[12pt]\label{bch4}
(M^{\pm}_{i})^\oplus = - M^{\pm}_{i} , \qquad (M^{\pm}_2)^\oplus  = M^{\pm }_2 \qquad 
\qquad \hbox{Kleinian case}
\\[12pt]\label{bch4x}
(M^{+}_r)^{\tilde{\oplus}} = M^{+}_r ,  \qquad
(M^{-}_i)^{\tilde{\oplus}} = - M^{-}_i ,\quad (M^{-}_2)^{\tilde\oplus}  = M^{-}_2  \  \
    \qquad \hbox{quaternionic case}                                                                     
\end{eqnarray}

Further we shall consider the supersymmetric N=1,2 extensions of 
 the Euclidean and Poincar\'{e} 
  real forms  (\ref{bch2}--\ref{bch4}).
The complexifications of real  Poincar\'{e},  Euclidean and Kleinian algebras are equvalent, and one can  consider as  well in place of reality constraints  (\ref{2a4})--(\ref{2a6})
  the real forms of complexified Poincar\'{e} or Kleinian algebras
 in order to provide the real Poincar\'{e}, Euclidean and Kleinian algebras.


It is quite useful to work further with Lorentzian (Poincar\'{e}) canonical basis which is obtained after realification of the Cartan-Chevaley basis of  $\mathfrak{sl}(2,\mathbb{C})$ \footnote{
Let $(h, e_\pm)$ be the Cartan-Weyl basis of $\mathfrak{sl}(2, \mathbb{C})$ with the commutation relations in the first line of (\ref{bor4.2}). 
   Setting $h':=  i h, e'_\pm :=  i e_\pm$ we obtain all commutation relations (\ref{bor4.2}). 
	The real Lie algebra $\mathfrak{sl}(2,\mathbb{C})_\mathbb{R}$ generated by the elements $(h, e_\pm, h', e'_\pm)$
with the defining relations (\ref{bor4.1}) 
 is called realification of $\mathfrak{sl}(2,\mathbb{C})$.}. 
    In such a basis Lorentz generators   are defined as follows (see \cite{T3,BLMT})
\begin{eqnarray}\label{bor4.1}
\begin{array}{rcccl}
h&\!\!=&\!\!-iN_3\ ,\qquad e_{\pm}&\!\!=&\!\!-i(N_1\mp M_2)\, ,
\\[6pt]
h'&\!\!=&\!\!iM_3\ ,\qquad\;\; e'_{\pm}&\!\!=&\!\!i(M_1\pm  N_2)\, .
\end{array}
\end{eqnarray}
One obtains the following description of $D=4$ Lorentz algebra
\begin{eqnarray}\label{bor4.2}
\begin{array}{rcccl}
[h,\,e_{\pm}^{}]\!\!&=\!\!&\pm e_{\pm}^{},\qquad [e_{+}^{},\,e_{-}^{}]\!\!&=\!\!&2h,
\\[6pt]
[h,\,e'_{\pm}]\!\!&=\!\!&\pm e'_{\pm}~,\qquad[h',\,e_{\pm}]\!\!&=\!\!&\pm
e'_{\pm},\qquad [e_{\pm}^{},\,e'_{\mp}]\;=\;\pm2h',
\\[6pt]
[h',\,e'_{\pm}]\!\!&=\!\!&\mp e_{\pm}^{},\qquad [e'_{+},\,e'_{-}]\!\!&=\!\!&-2h \, .
\end{array}
\end{eqnarray}
The fourmomenta generators with the components
\begin{equation}\label{bor4.3}
P_\mu= (P_1,\;\;P_2,\;\;P_{\pm}\;=\;P_{4}\pm P_{3})
\end{equation}
extend (\ref{bor4.2}) 
  to the real Poincar\'{e} algebra as follows
\begin{eqnarray}\label{bor4.4}
\begin{array}{rcccccccl}
[h,\,P_{\pm}]&\!\!=&\!\!\pm P_{\pm},\qquad[h,\,P_{i}]&\!\!=&\!\!0\quad\;\;(i=1,2),\quad&&\qquad &&
\\[5pt]
[e_{\pm},\,P_{\pm}]&\!\!=&\!\!0,\qquad\;\;\;[e_{\pm},\,P_{\mp}]&\!\!=&\!\!2P_{1},\quad\;\; [e_{\pm},\,P_{1}]&\!\!=&\!\!P_{\pm},\quad\;\;[e_{\pm},\,P_{2}]&\!\!=&\!\!0 \, ,
\end{array}
\end{eqnarray}
\begin{eqnarray}\label{bor4.4'}
\begin{array}{rcccccccl}
[h',\,P_{\pm}]&\!\!=&\!\!0,\qquad\; [h',\,P_{1}]&\!\!=&\!\!-P_{2},\qquad [h',\,P_{2}]&\!\!=&\!\!P_{1},\qquad\qquad\quad &&
\\[5pt]
[e_{\pm}',\,P_{\pm}]&\!\!=&\!\!0,\qquad[e_{\pm}',\,P_{\mp}]&\!\!=&\!\!\mp2P_{2},\quad \;\;[e_{\pm}',\,P_{1}]&\!\!=&\!\!0,\qquad[e_{\pm}',\,P_{2}]&\!\!=&\!\!\mp P_{\pm}\, .
\end{array}
\end{eqnarray}
Three reality conditions imposed on these canonical generators are now
\begin{eqnarray}\label{v2.2}
x_A^\dagger&\!\!=&\!\!-x_A\quad\hbox{for}\;\; x_A\in(h,\;e_{\pm},\;h',\;e'_{\pm},\;P_{\pm},\;P_1,\;P_2)\quad\hbox{( Poincar\'{e} case )}
\end{eqnarray}
\begin{eqnarray}\label{ch3}
\begin{array}{rcl}
h^{\ddagger}&\!\!=&\!\!h,\qquad\; e_{\pm}^{\ddagger}\;=\;e_{\mp},\quad\;\;{h'}^\ddagger\;=\;-h',\quad {e'}_{\pm}^\ddagger\;=\;-{e'}_{\mp},
\\[5pt]
P_{\pm}^{\ddagger}&\!\!=&\!\!-P_{\mp},\quad P_{i}^{\ddagger}\;=\;P_{i}\quad (i=1,2)
\qquad\qquad\qquad\hbox{( Euclidean case )}
\end{array}
\end{eqnarray}
\begin{eqnarray}\label{ch3c}
\begin{array}{rcl}
h^{\oplus}&\!\!=&\!\!-h,\qquad\; e_{\pm}^{\oplus}\;=\;-e_{\pm},\quad\;\;{h'}^\oplus\;=\;h',\quad {e'}_{\pm}^\oplus\;=\;{e'}_{\pm},
\\[5pt]
P_{\pm}^{\oplus}&\!\!=&\!\!-P_{\pm},\quad
 \tilde{P}{}_{\pm}^{\oplus}\;=-\; \tilde{P}{}_{\pm}^\oplus ,
\quad 
\tilde{P}{}_{\pm} \;=\; P_{1} \pm i P_2
\qquad\quad\hbox{( Kleinian case )}
\end{array}
\end{eqnarray}

\subsection{Complex $D=4\  N=1$ Euclidean superalgebra}

In this paper we shall consider the superalgebra generators in purely algebraic way, without reference to concrete
realizations. In this subsection we shall recall the complex $D=4$ $N=1$ Euclidean superalgebra (see e.g. \cite{LukierskiNo,Luk,FW}) describing simple ($N=1$) supersymmetrization of $\mathcal{\epsilon}(4;\mathbb{C})$ (inhomogeneous $\mathfrak{o}(4;\mathbb{C})$) complex algebra. Such superalgebra is obtained by adding to the generator of complex $D=4$ Euclidean algebra $\epsilon(4;C)=\mathfrak{o}(4;C)\ltimes
\mathbf{T}^{C}_4$
 four independent complex supercharges $Q_\alpha, \bar Q_{\dot{\alpha}}$ transforming as
 fundamental representations under ``left'' and ``right''  internal symmetry groups
 $SL_+(2;C)$ and $SL_-(2;C)$~\footnote{We recall that $O(4;C)=O_+(3;C) \otimes O_- (3;C)$ has a spinorial covering $SL_+(2;C) \otimes SL_- (2;C)$. The $SL_+ (2;C)$ spinors have undotted spinorial indices, and dotted indices characterize $SL_- (2;C)$ spinors. These two groups are also called chiral (left) and antichiral (right) projections of
 $\mathfrak{o}(4;C)$ group.}.
  The supercharges $Q_\alpha, Q_{\dot{\alpha}}$  extend the $D=4$ complex Euclidean 
	 algebra (\ref{2a3}) 
	 by the following algebraic relations
  \begin{eqnarray}{}\label{2s210}
  &\left\{  Q_\alpha , \bar Q_{\dot{\beta}}\right\}
  = 
 2(\sigma^E_\mu)_{\alpha \dot{\beta}} {\mathcal P}^\mu
  \qquad \quad
   \left\{  Q_\alpha , Q_{{\beta}}\right\}
   =    \left\{ \bar Q_{\dot{\alpha}} , \bar Q_{\dot{\beta}}\right\} = 0
	\\[-6pt] \label{2s211} 
  &\left[
  {\mathcal L}_{\mu\nu}, Q_\alpha \right]   = 
  -(\sigma^E_{\ \mu\nu})_{\alpha}^{\  \ \beta} \, Q_\beta
  \qquad \qquad \qquad
  \left[
  {\mathcal L}_{\mu\nu}, \bar Q_{\dot{\alpha}} \right]   = 
  \bar Q_{\dot{\beta}}({\tilde{\sigma}}^E_{\ \mu\nu})^{\dot{\beta}}_{\ \dot{\alpha}}
	\\ 
  & \qquad \quad 
	\left[
  {\mathcal P}_\mu , Q_\alpha
  \right]  = 
  \left[
  {\mathcal P}_\mu, \bar Q_{\dot{\alpha}}
  \right] = 0  \, ,
	 \label{2s212}
  \end{eqnarray}
  where the Euclidean sigma matrices $\sigma^E_\mu$ are expressed by standard Pauli matrices $\sigma_r(r=1,2,3)$ as follows
  \begin{equation}\label{2s213}
  (\sigma^E_\mu)_{\alpha \dot\beta}  =
   ( (\sigma_{\, r})_{\alpha {\dot{\beta}}}, i(I_{\, 2})_{\alpha {\dot{\beta}}})
   \qquad r=1,2,3
  \end{equation}
  and satisfy the known reality  conditions under Hermitean matrix conjugation
  \begin{equation}\label{2s214}
  (\sigma^E_{ r})^+   = \sigma^E_{ r} \qquad  (\sigma^E_{\, 0})^+ = - \sigma_{\, 0}\, .
  \end{equation}
  The matrices $\sigma^E_{\mu\nu}$ and $\tilde{\sigma}{}^E_{\mu\nu}$ describe the following realizations of
  the pair of commuting $sl_\pm (2;C)$ algebras

  \begin{eqnarray}\label{2s215}
  sl_{\, +}(2;C): \qquad (\sigma^E_{\, \mu\nu})_{\alpha}^{\ \ \beta}
  & = &
  (\sigma^{\ \ E}_{\, rs} = {1\over 2} \epsilon_{rst} \sigma_t,\ \ \sigma^{\ \ E}_{\, 0 r} =
  \frac{1}{2} \sigma_{\,r})\, ,
  \nonumber
  \\
    sl_{\, -}(2;C): \qquad ({\tilde{\sigma}}^{\ \ E}_{\, \mu\nu})_{\dot{\alpha}}^{\ \ \dot{\beta}}
  & = &
  (\tilde{\sigma}^{\ \ E}_{\, rs} = \frac{1}{2} \epsilon_{rst} \sigma_t,\ \  {\tilde{\sigma}}^{\ E}_{\, 0 r} =
  - \frac{1}{2} \sigma_{\,{r}}) \, .
  \end{eqnarray}
  The complexified Euclidean fourmomenta ${\mathcal P}_{\alpha \dot{\beta}} \equiv (\sigma^{\ E}_\mu)_{\alpha \dot{\beta}} \, {\mathcal P}^\mu$
  transform under the complex Euclidean rotations $A^\mu_\nu\in O(4;C)$ described by the product of two commuting
  Lorentz groups $SL_\pm(2;C)$  as follows ($S_\alpha^{\ \ \beta} \in SL_{\, +} (2;C)$,
  $\tilde{S}_{\dot{\alpha}}^{\ \ \dot{\beta}} \in S_{-} (2;C)$)

  \begin{equation}\label{2s216}
  {\mathcal P}'_{\, \alpha \dot{\beta}} =
   S_\alpha ^{\ \ \gamma} \, \tilde{S}_{\, \dot{\beta}}{}^{\ \dot{\delta}}
    {\mathcal P}_{\gamma \dot{\delta}}\ \  \longleftrightarrow  \ \
     {\mathcal P}'_\mu = A_\mu^\nu {\mathcal P}_\nu
  \end{equation}
where $\mathcal{P}_\mu={1\over 2}\sigma_\mu^{\alpha\dot{\beta}}\mathcal{P}_{\alpha\dot{\beta}}$  and
$A^\mu_\nu=\sigma_\nu^{\alpha\dot{\beta}}\sigma^\mu_{\gamma\dot{\delta}}S_\alpha^\gamma
\tilde S^{\dot{\delta}}_{\dot{\beta}}$ .

  Finally one can check that the relations (\ref{2s210}--\ref{2s212}) 
	 are invariant under the complex rescaling transformation (c is a complex number)
  \begin{equation}\label{2s221}
  Q'_\alpha = c Q_\alpha \qquad \quad \bar Q'_{\dot{\alpha}} = c^{-1} \, \bar Q_{\dot\alpha}\, .
  \end{equation}
  The rescaling (\ref{2s221}) 
	 is described by $GL(1;C)=U(1)\times R$ Abelian group and represents the  one-parameter complex internal symmetries of $D=4$ simple complex Euclidean superalgebra. The N=1 internal $gl(1;C)$ generator $T$ satisfies the algebraic relation
  \begin{equation}\label{2s222b}
   [T, Q_\alpha]=Q_\alpha
  \qquad
   [T, \bar Q_{\dot\alpha}]=-\bar Q_{\dot\alpha} \, .
  \end{equation}

\subsection{Real and pseudoreal forms of $D=4$ $N=1$ complex Euclidean superalgebra.}

\subsubsection{$D=4$ $N=1$ pseudoreal Euclidean superalgebra and its pseudoreal forms.} The reality conditions (\ref{2a5}) 
  defining  $D=4$ Euclidean space-time algebra ${\mathfrak{o}}(4;\mathbb{R})\ltimes\mathbf{T}_4$
 can be extended to the sector of Euclidean spinorial supercharges  if we use the 
    corresponding 
    spinorial  covering  group $\overline{{{O}}(4;\mathbb{R})}= SU_+(2)\oplus  SU_-(2)$    which 		requires a pair of independent  two-component complex ${SU}_{}(2)$  spinors.
This property of doubling of $D=4$ Euclidean spinorial components in comparison with standard relativistic $D=4$ case leads to the known conclusion that contrary to the Poincar\'{e} case the four-component real (Majorana) Euclidean spinors do not exist. In the algebraic framework one can however extend in odd supercharges sector the conjugation (\ref{2a5}) 
as the pseudoconjugation
(see Appendix A, (\ref{3s3b}b)) 
  which should be consistent with $N=1$  complex Euclidean algebra (\ref{2s210}--\ref{2s212}) 
		under the  assumption that the generators ${\mathcal P}_\mu$, $\mathcal{L}_{\mu\nu}$ are real Euclidean, i.e. satisfy  the reality conditions (\ref{2a5}). 
	 The $N=1$ pseudoconjugation of Euclidean supercharges  is an involution of fourth order
 (see also  (\ref{3s3b}b)) satisfying the relation
\begin{equation}\label{3s312before}
(Q^{\ddagger}_{\alpha})^\ddagger\;\;=\;\;\ -\ Q_{\alpha},\qquad(\bar{Q}^{\ddagger}_{\dot\alpha})^\ddagger\;\;=\;\; -\ \bar{Q}_{\dot\alpha}\
\end{equation}
and it look as follows \cite{Castro,Ferrara,Buch}
\begin{equation}\label{3s312}
Q^{\ddagger}_{\alpha}\;\;=\;\;\epsilon_{\alpha\beta}Q_{\beta},\qquad\bar{Q}^{\ddagger}_{\dot\alpha}\;\;=\;\; \eta\epsilon_{\dot{\alpha}\dot{\beta}}\bar{Q}_{\dot\beta}\ , \qquad \eta=\pm 1 \, .
\end{equation}
It can be shown that the map $Q_\alpha\rightarrow -\epsilon_{\alpha\beta}Q^\ddagger_\beta,  \bar Q_{\dot\alpha}\rightarrow -\eta\epsilon_{\dot\alpha\dot\beta}Q^\ddagger_{\dot\beta}$  leaves the  
superalgebra (\ref{2s210}--\ref{2s212}) 
 invariant provided that we choose the parameter $q$ (see Appendix,  A (\ref{3s31})) 
  in accordance with the relation
  \begin{equation}\label{x2.36}
	\eta =(-1)^{q+1} \, .
	\end{equation}
We obtain the following two cases:

$i) \ \eta=-1,  q=0$ (standard choice)

In this case the product of odd (fermionic operators $f_1, f_2$ which are odd powers 
 of supercharges are conjugated as follows

\begin{equation}\label{x2.37}
(f_1f_2)^\ddagger = f_2^\ddagger f_1^\ddagger
\end{equation}
and leads to standard supersymmetry scheme with pseudoconjugation $\ddagger$ which can be represented on complex Grassmannian variables
 $\theta_\alpha = \theta^1_\alpha + i \theta^2_\alpha$ 
   as the complex 
    conjugation $\theta_\alpha \to \theta^\star_\alpha = \theta^1_\alpha - i \theta^2_\alpha$.
  It should be added that all applications of pseudoconjugations to the description of real
	$N=1$  Euclidean SUSY use the case  $i)$  (see e.g. \cite{IvanovZu}).

$ii)\  \eta=1, q=1$ (exotic choice)

In this case the product of odd (fermionic) operators $f_1, f_2$ which are odd powers of supercharges and Grassmann variables  are conjugated as follows 

\begin{equation}\label{x2.39} 
(f_1f_2)^\ddagger=-f_2^\ddagger f_1^\ddagger \, .
\end{equation}

Such choice leads to nonstandard supersymmetry scheme, which can be realized in complex superspace described by odd Grassmann variables $\tilde\theta_A$,  ${\bar{\tilde{\theta}}}_A$ with the products transformating under conjugation
	  in the nonstandard way

\begin{equation}\label{xy2.39} 
(\tilde\theta_A \, \tilde\theta_B)^\ddagger=
-\tilde\theta_B^\ddagger\,  \tilde\theta_A^\ddagger \, ,
\qquad 
(\bar{\tilde\theta}_A \bar{\tilde\theta}_B)^\ddagger =
-(\bar{\tilde\theta})_B^\ddagger \, (\bar{\tilde\theta})_A^\ddagger \, .
\end{equation}

Such type of Grassmann variables was considered as mathematically consistent choice in \cite{Scheunert, BerVNT, YuManin} but it has not been applied
 in the literature to describe the physical supersymmetric systems, therefore exotic.

In the  exotic case the conjugation $A\mapsto A^{\ddagger}$ (antilinear antiautomorphism of second order) in bosonic sector of the Euclidean superalgebra defining the reality condition
 (\ref{2a5}) 
  is lifted in fermionic sector of supercharges to the antilinear antiautomorphism of fourth order  defining  pseudoconjugation.

One can show that for both values $\eta=\pm 1$ in the scaling transformations  (\ref{2s221}) 
 the parameter
$c$ should be  real,  what means that the invariance of  (\ref{2s222b})  under the 
 pseudoconjugation (\ref{3s312}) 
is valid if $T^\ddagger=-T$, i.e.  $N=1$ internal symmetry $GL(1;\mathbb{C})=U(1)\oplus O(1,1)$ is reduced to $GL(1;\mathbb{R})=O(1,1)$ for pseudoreal $N=1$ Euclidean superalgebra.

\subsubsection{$D=4$ $N=1$ real Poincar\'{e} superalgebra}

$N=1$ Poincar\'{e} superalgebra is obtained from the relations 
(\ref{2s210}--\ref{2s212})  
  after imposing the following extension of the Poincar\'{e}
 reality condition  (\ref{2a4}) 
  to the supercharges sector
\begin{equation}\label{sp2_0}
(Q_\alpha)^{\dagger}\;\;=\;\;\bar{Q}_{\dot{\alpha}},\qquad(\bar{Q}_{\dot{\beta}})^{\dagger}\;\;=\;\;\eta Q^{}_{\beta}\qquad \eta=\pm 1 \, .
\end{equation}
 If $\eta=1, \ q=0$ the formula 
 (\ref{sp2_0}) 
 describe conjugation and lead to standard $N=1$ real Poincar\'{e} supersymmetry.
  If $\eta=-1$ we get  $q=1$  
   and relations  (\ref{sp2_0}) 
 describe pseudoconjugation  (pseudoreality condition) defining exotic supersymmetry  with Grassmann variables satisfying relations 
  (\ref{x2.39}). 
		
After introducing $P_i = \mathcal{P}_i$,  $P_0= i \mathcal{P}_0$  
  (see \ref{2a2}) 
  one gets   from  (\ref{2s210}--\ref{2s212}) the well-known  N=1 Poincar\'{e} superalgebra with
the pair of supercharges represented by two-component complex Weyl spinors
\begin{eqnarray}\label{sp4_0}
\begin{array}{rcccl}
\{Q^{}_{\alpha},\,\bar{Q}_{\dot{\beta}}\}\!\!&=\!\!&2\,(\sigma_{\mu})_{\alpha{\dot{\beta}}}
P^{\mu},\qquad  [P_{\mu}^{},\,Q^{a}_{\alpha}]\!\!&=\!\!&[P_\mu^{},\,\bar{Q}_{\dot{\alpha}a}]\;\;=\;\;0 \, ,
\\[7pt] 
\{Q^{}_{\alpha},\,Q^{}_{\beta}\}\!\!&=\!\!&0,\qquad\qquad\qquad
\{\bar{Q}_{\dot{\alpha}},\,\bar{Q}_{\dot{\beta}}\}\!\!&=\!\!& 0  \, ,
\\[7pt]  
[L_{\mu\nu},\,Q^{}_{\alpha}]\!\!& =\!\!&-\displaystyle{(\sigma_{\mu\nu})_{\alpha}}^{\beta}Q^{a}_{\beta},\qquad
[L_{\mu\nu},\,\bar{Q}_{\dot{\alpha}a}]\!\!& =\!\!&\displaystyle\bar{Q}_{\dot{\beta}a}{(\tilde{\sigma}_{\mu\nu})}_{\dot{\alpha}}^{\dot{\beta}} \,  , 
 \end{array}
\end{eqnarray}
where the Lorentzian $\sigma-$matrices are defined as follows
\begin{eqnarray}\label{sp3_0}
\begin{array}{rcl}
\sigma_{\mu} &= & (\sigma_i\;\;=\;\;\sigma^{E}_{i},\;\;\;\sigma_{4}\;=  i \sigma^{E}_{4})\, ,
\\[7pt] 
\sigma_{\mu\nu} & = & (\sigma_{ij}\;=\;\sigma^{E}_{ij},\;\;\sigma_{4i}\;=\;- i \sigma^{E}_{4i})]\, , 
\\[7pt] 
\tilde{\sigma}_{\mu\nu} & = & (\tilde{\sigma}_{ij}\;=\;\widetilde\sigma^{E}_{ij},\;\;\tilde\sigma_{4i}\;=\;-  i \tilde\sigma^{E}_{4i}) \, .
\end{array}
\end{eqnarray}
It is easy to see that for $\eta=\pm 1$ the pair of relations (\ref{2s222b}) 
   is invariant under the conjugation  (\ref{sp2_0}) 
	 if $T^\dagger=T$, i.e. the internal symmetry of  $N=1$ Poincar\'{e}  supersymmetry is described by the restriction of $GL(1;\mathbb{C})$ to $U(1)$.

\subsubsection{$D=4$ \ $N=1$ real  Kleinian superalgebra}

The $N=1$ Kleinian $\mathfrak{o}(2,2)$ superalgebra is obtained after the following extension of the reality  condition  (\ref{2a6}) 
  to the supercharges sector
\begin{equation}\label{sp2_0k}
(Q_\alpha)^{\oplus}\;\;=\;\;Q^{}_{\alpha},\qquad(\bar{Q}_{\dot{\beta}})^{\oplus}\;\;=\;\;\eta\bar{Q}_{\dot{\beta}}\qquad \eta=\pm 1 \, ,
\end{equation}
i.e. the supercharges (\ref{sp2_0k})
  form a pair of respectively $\mathfrak{o}_+(2,1)=\mathfrak{sl}_+(2;\mathbb{R})$ and
$\mathfrak{o}_-(2,1)=\mathfrak{sl}_-(2;\mathbb{R})$ real spinors.
 Imposing on  complex superalgebra (\ref{2s210}--\ref{2s212}) 
  the reality condition  (\ref{sp2_0k}) one gets the following real superalgebra
\begin{equation}\label{kk1}
    \{Q^{}_{\alpha},\,\bar{Q}_{\dot{\beta}}\}\!\!=\!\!2\,({\tilde{\sigma}}_{\mu})_{\alpha{\dot{\beta}}}\tilde P^\mu \, ,
\end{equation}
where $\tilde P_\mu\tilde P^\mu=\tilde P^2_1-\tilde P^2_2+\tilde P^2_3-\tilde P^2_4$ and
\begin{equation}\label{kk2}
    \tilde P_\mu = (\mathcal{P}_1,  i \mathcal{P}_2, \mathcal{P}_3,  i \mathcal{P}_4)
\end{equation}
with $\tilde\sigma_\mu$ describing $\mathfrak{o}(2, 2)$ real $\sigma-$matrices
\begin{equation}\label{ykk2}
    \tilde \sigma_\mu = (\sigma_1,  i \sigma_2, \sigma_3, -I_2) \, .
\end{equation}

The inhomogeneous $\mathfrak{o}(2,2)$ algebra can be described more conveniently if we use the formulation of complex Euclidean algebra $\mathcal{\epsilon}(4;\mathbb{C})$  using formulae 
 (\ref{bch1}--\ref{bch1bis}).  
   In  Kleinian case the "complex"
generators $M^\pm_r$ become the set of real generators (i.e. generating real Lie superalgebra with real structure constants)

\begin{eqnarray}\label{kk3}
\begin{array}{rcccl}
[M^{(+)}_{r},\,Q^a_{\alpha}]\!\!&=\!\!&-\frac{1}{2}{(\tilde{\sigma}_{r})_{\alpha}}^{\beta}Q^{a}_{\beta},\qquad\;
[ M^{(-)}_{r},\,Q^{a}_{\alpha}]\!\!&=\!\!&0,
\\[7pt]
[M^{(+)}_{r},\,\bar Q_{\dot{\alpha}a}]\!\!&=\!\!&0,\qquad\qquad\qquad\quad
[M^{(-)}_{r},\,\bar Q_{\dot{\alpha}a}]\!\!&=\!\!&\frac{1}{2}\bar
Q_{\dot{\beta}a}{(\tilde{\sigma}_{r})^{\dot{\beta}}}_{\dot{\alpha}} \, .
\end{array}                                                             \end{eqnarray}
We add that similarly as in Poincar\'{e} case, the choice $\eta=-1,\ q=0$ leads to standard $N=1$ Kleinian supersymmetry
  with (\ref{sp2_0k}) 
	 describing conjugation $\oplus$ (reality conditions), 
 and if  $\eta=1,\ q=1$ one gets (see also Sec. 2.3.1) exotic supersymmetry with pseudoconjugation implying the exotic antiautomorphism  which leads to relations (\ref{x2.39}). 
 Finally for both values of $\eta$ the relations  (\ref{2s222b}) 
 are consistent with the reality condition 
  (\ref{sp2_0k}) 
 if $T^\oplus=-T$, i.e. real $N=1$ Kleinian symmetry is endowed with $GL(1;\mathbb{R})\equiv O(1,1)$ internal symmetry. 

\section{Complex $D=4$  $N=2$ Euclidean supersalgebra  and its Euclidean, Poincar\'{e} and Kleinian (pseudo) real forms}
\sec
\subsection{Complex $D=4$ $N\negthinspace=\negthinspace2$ Euclidean   superalgebra}
In this subsection we shall describe $N=2$ superization of complex inhomogeneous $\mathfrak{o}(4;\mathbb{C})$ algebra.  Basic N=1 relations (\ref{2s210}) 
 are extended to $N=2$  as follows ($a,b..=1, 2$):
  \begin{eqnarray}{}\label{2s2101}
  \left\{  Q^a_\alpha , \bar Q_{\dot{\beta\,b}}\right\}
 & = &
 2\, \delta^a_b\, (\sigma^E_\mu)_{\alpha \dot{\beta}} {\mathcal P}^\mu \, ,
  \\[-6pt] \nonumber
  \\[-6pt] \label{2s2111} 
   \left\{  Q^a_\alpha , Q^b_{{\beta}}\right\}=\epsilon_{\alpha\beta}\epsilon^{ab}Z\, ,
  &   &   \left\{ \bar Q_{\dot{\alpha}a} , \bar Q_{\dot{\beta}b}\right\} = \epsilon_{\dot{\alpha}\dot{\beta}}\epsilon_{ab}\tilde Z \, ,
   \end{eqnarray}
where $Z,\tilde Z$ describe a pair of complex scalar central charges.
The relations (\ref{2s2101}) 
  are invariant under the internal symmetries $A\in GL(2, C)$, where
  \begin{equation}\label{2s223b}
  Q^{a}{}_\alpha \rightarrow (A)^a {}_b\,Q^b_\alpha
  \qquad
  \bar Q_{{\dot{\alpha}} a} \rightarrow \bar Q_{{\dot{\alpha}} b}\,(A^{-1})^b_{\ \ a} \, .
  \end{equation}
The presence of central charges breaks only internal symmetry ${GL}(2;\mathbb{C})$
to ${SL}(2;\mathbb{C})$, what follows from the relation $A\epsilon A^T =\epsilon$ valid for complex $2\times 2$ matrices $A \in SL(2;\mathbb{C})$ due to the equivalence $SL(2;\mathbb{C})\sim Sp(2;\mathbb{C})$. The $\mathfrak{gl}(2;\mathbb{C})$ generators ${T_{i}}^{j}$ \ ($i,j=1,2$)
\begin{eqnarray}\label{2s227x}
[{T_{i}}^{j},\,{T_{k}}^{l}]\!\!&=\!\!& i\, ({\delta^{j}}_{k}{T_{i}}^{l} -{\delta^{l}}_{i}{T_{c}}^{k})
\end{eqnarray}
 are restricted to $\mathfrak{sl}(2;\mathbb{C})$ by the condition $T_0\equiv {T_{1}}^{1}+{T_{2}}^{2}=0$.

One can describe $N=2$ internal symmetries algebra $\mathfrak{gl}(2;\mathbb{C})$ in convenient way by  replacing four generators $T_i^j$ 
by generators $T_A (A=0,1,2,3$) adjusted to further description of $U(2)$ internal symmetry
 \begin{equation}\label{x1}
 T_A= { i \over 2}(\sigma_A)^i_j T^j_i \, ,
 \end{equation}
 where $\sigma_A = (\sigma_r, \sigma_0=I_2)$; $\sigma_r$ are three $2x2$ Hermitean Pauli matrices.
 The $\mathfrak{gl}(2;\mathbb{C})$ covariance relation of $N=2$ supercharges looks as follows  
 \begin{eqnarray}\label{x1a}
[{T_{A}},\,Q_{\alpha}^{a}]\!\! =\!\! -\frac{1}{2}{({\sigma_{A}}^{})^{a}}_{b}Q^{b}_\alpha \, , 
\qquad
[{T_{A}},\,\bar Q_{\dot{\alpha}a}]\!\!=\!\!\frac{1}{2}\bar Q_{\dot{\alpha}b}{({\sigma_{A}}^{})^{b}}_{a} \, ,
\end{eqnarray}
where the $2x2$ matrices $(t_A)_a ^b =   
{1\over 2}(\sigma_A)_a ^b$ provide the fundamental realizations of $\mathfrak{gl}(2,C)$  
 generators (\ref{x1})\footnote{It corresponds to the realization $(T^j_i)^b_a= i \delta^b_i\delta^j_a$ of the generators in the  relations (\ref{2s227x}).}. 
 Using known relation ($r,s,t=1,2,3$)
 \begin{equation}\label{x1b}
 \sigma_r\sigma_s=\delta_{rs} + i \epsilon_{rst}\sigma_t
 \end{equation}
 one gets from  (\ref{2s227x}) 
  and (\ref{x1}) 
	 the $N=2$ internal  symmetry algebra $\mathfrak{gl}(2)$
 \begin{equation}\label{xx}
  [T_r, T_s]= i \epsilon_{rst}T_t  \qquad [T_0, T_r]=0 \, .
 \end{equation}
 Three generators $T_r$ span $\mathfrak{sl}(2;\mathbb{C})$ algebra which is preserved even in the case of nonvanishing central charges $Z, \tilde Z$; the Abelian generator 
 $T_0$  
  describing coset ${GL}(2;\mathbb{C})\over  {SL}(2;\mathbb{C})$
  is broken 
 in the presence of central charges.
 
If we consider the chiral projections of Euclidean $\mathfrak{o}(4;\mathbb{C})$ 
generators (see (\ref{bch1})--(\ref{bch1bis})) 
  one gets from  (\ref{x1a}) 
		 the commutators exposing the chiral 
(antichiral) nature
 of supercharges $Q^i_\alpha$ $(\bar{Q}_{\dot{\alpha}i})$ 
\begin{eqnarray}\label{2s230}
\begin{array}{rcccl}
[M^{(+)}_{r},\,Q^a_{\alpha}]\!\!&=\!\!&-\frac{1}{2}{(\sigma_{r})_{\alpha}}^{\beta}Q^{a}_{\beta}\, ,\qquad\;
[ M^{(-)}_{r},\,Q^{a}_{\alpha}]\!\!&=\!\!&0\, ,
\\[7pt]
[M^{(+)}_{r},\,\bar Q_{\dot{\alpha}a}]\!\!&=\!\!&0\, ,\qquad\qquad\qquad\quad
[M^{(-)}_{r},\,\bar Q_{\dot{\alpha}a}]\!\!&=\!\!&\frac{1}{2}\bar Q_{\dot{\beta}a}{(\sigma_{r})^{\dot{\beta}}}_{\dot{\alpha}}\, .
\end{array}
\end{eqnarray}
The vanishing commutators in (\ref{2s230}) 
  illustrate that the supercharges $Q^a_\alpha$ are left-handed (chiral) and the supercharges $\bar Q_{\dot{\alpha}a}$ are right-handed (antichiral).\\

\subsection{$N=2$ real Poincar\'{e} superalgebras with central charges}

The reality conditions for supercharges  can take the form (see e.g. \cite{Haag,Frappat}):
\begin{equation}\label{sp2}
(Q^a_\alpha)^{\dagger}\;\;=\;\;\bar{Q}_{\dot{\alpha}a}\, ,
\qquad(\bar{Q}_{\dot{\beta}b})^{\dagger}\;\;=\;\; \eta \, Q^{b}_{\beta} 
\qquad \eta = \pm 1 \, .
\end{equation}

The reality condition for well-known $N=2$ real Poincar\'{e}  superalgebra with central charges \cite{Haag} is obtained
 if we put  $q=0$ (see (A.1)) and  $\eta=1$. 
 In such a case we impose on the complex generators $\{{\mathcal{L}}_{\mu\nu},\;{\mathcal{P}}_\mu,\;Q^i_\alpha,
\;\bar{Q}_{\dot{\beta}j},\; T_{r}\;=\;\frac{1}{2}{(\sigma_{r})^{i}}_{j}{T_{i}}^{j},\;Z_{1},\;Z_{2} ;\; i\;=\;1,2;\;r\;=\;1,2,3\}$ of centrally extended $N=2$ complex Euclidean  superalgebra 
the reality constraints which extend consistently the conjugation (\ref{2a3}) 
  in bosonic
sector to odd superalgebra generators.

In particular in the representation which permits the Hermitean conjugation of supercharges the 
 conjugation (\ref{sp2}) 
  can be seen as  Hermitean conjugation. Further the reality constraints on the internal symmetry generators $T_A$ (see (\ref{x1}) 
	 and (\ref{xx})) 
	 and the central charges ($Z_1,Z_2$),  which are consistent with the 
	relations (\ref{2s2101}--\ref{2s2111}),  (\ref{x1a}) 
 and (\ref{xx}) 
  are the following
\begin{equation}\label{sp3}
T_{r}^{\dagger}\;\;=\;\;T_{r}\, ,\qquad T_0^\dagger =-T_0\, ,\qquad Z_{1}^{\dagger}\;\;=\;\;Z_{2},\qquad Z_{2}^{\dagger}\;\;=\;\;Z_{1}\, ,
\end{equation}
where the generator $T_0$  describes internal symmetry only in the case when central charges vanish. If the central charges are not vanishing
from first set of the relations (\ref{sp3}) 
  one can see that the algebra $\mathfrak{sl}(2;\mathbb{C})$   is constrained to $\mathfrak{su}(2)$ algebra.
If we use  the formulae (\ref{sp3_0}) for 
  Minkowskian $\sigma$-matrices, it follows
from (\ref{2a2}) 
  and (\ref{2s213}) 
	 that $\sigma_\mu P^\mu=\sigma^E_\mu P^{E\mu}$ and we get the following real $N=2$ Poincar\'{e} superalgebra with one complex central charge $Z$:
\begin{eqnarray}\label{sp4}
\begin{array}{rcccl}
\{Q^{a}_{\alpha},\,\bar{Q}_{\dot{\beta}b}\}\!\!&=\!\!&2\delta^{a}_{b}(\sigma_{\mu})_{\alpha{\dot{\beta}}}
P^{\mu},\qquad\qquad\qquad && 
\\[7pt] 
\{Q^{a}_{\alpha},\,Q^{b}_{\beta}\}\!\!&=\!\!&\epsilon^{ab}\epsilon_{\alpha\beta}Z,\qquad\qquad
\{\bar{Q}_{\dot{\alpha}a},\,\bar{Q}_{\dot{\beta}b}\}\!\!&=\!\!&\epsilon_{ab}\epsilon_{\dot{\alpha}\dot{\beta}}\bar{Z}, 
\\[7pt]  
[L_{\mu\nu},\,Q^{a}_{\alpha}]\!\!& =\!\!&-\displaystyle{(\sigma_{\mu\nu})_{\alpha}}^{\beta}Q^{a}_{\beta},\qquad
[L_{\mu\nu},\,\bar{Q}_{\dot{\alpha}a}]\!\!& =\!\!&\displaystyle\bar{Q}_{\dot{\beta}a}{(\tilde{\sigma}_{\mu\nu})}_{\dot{\alpha}}^{\dot{\beta}}, 
\\[7pt]  
[T_{r},\,Q^{a}_{\alpha}]\!\!&=\!\!&-\displaystyle {(t_{r})^{a}}_{b}Q^{b}_{\alpha},\qquad\qquad
[T_{r},\,\bar{Q}_{\dot{\alpha}a}]\!\!&=\!\!&\displaystyle\bar{Q}_{\dot{\alpha}b}{(t_{r})^{b}}_{a},
\\[7pt] 
[P_{\mu}^{},\,Q^{a}_{\alpha}]\!\!&=\!\!&[P_\mu^{},\,\bar{Q}_{\dot{\alpha}a}]\;\;=\;\;0,\qquad\quad
[T_{r},\;T_{s}]\!\!&=\!\!& i \epsilon_{rsm}T_{m}\, . 
\end{array}
\end{eqnarray}
If $Z\neq0$ the $N=2$ internal symmetries ($R$-symmetries) are described by 
 $\mathfrak{su}(2)$ algebra, with the fundamental $2\times 2$ matrix realizations  
(see (\ref{2s210}) 
  and (\ref{sp4}) 
	  (the fourth line)) described by Hermitean Pauli matrices ($t_{r}\equiv\sigma_{r}$).
If $Z=0$ in relations (\ref{sp4}) 
    one can add fourth $R$-symmetry generator $T_0$ describing the extension of $\mathfrak{su}(2)$ to $\mathfrak{u}(2)$. In such a case the complex $GL(1;\mathbb{C})$ rescaling in complex N=2 Euclidean superalgebra generated by  $T_0$ 
(see (\ref{x1}) 
   and (\ref{xx})); 
   is restricted to $U(1)$ phase transformations
\begin{equation}\label{bor3.15}
{Q_{\alpha}^{l}}'\;\;=\;\;(\exp^{ i \xi})Q^{l}_{\alpha},\qquad
\bar{Q}_{\dot{\beta}i}'\;\;=\;\;(\exp^{- i \xi})\bar{Q}_{\dot{\beta}i} \, .
\end{equation}


If $\eta=-1$ the map (\ref{sp2}) 
  describes a
pseudoconjugation, but requires the exotic version ($q=1$) of formula (\ref{3s31}), 
   i.e. we get the following relation
 between $\eta$ in (\ref{sp2}) 
  and parameters $q$ (see (\ref{3s31})) 
\begin{equation}\label{xsp2a}
    \eta=(-1)^{q} \, .
\end{equation}

Further it can be shown that the antiautomorphism (\ref{sp2}) 
  leads to the following reality constraints on central supercharges $Z, \tilde Z$
\begin{equation}\label{xsp2b}
    Z^\dagger=-\eta\tilde Z\ ,\qquad\qquad \tilde Z^\dagger=-\eta Z \, .
\end{equation}
The antiautomorphism of complex $N=2$ relations (see (\ref{2s2101})--(\ref{xx})) 
 leads for both cases $\eta=\pm 1$ to the same 
restriction of internal  $gl(2,C)$ to its subgroup: $SU(2)$ in the presence of central charge ($Z\neq 0$)
and $U(2)$ if $Z=0$. We see therefore that we obtain the same internal symmetry sectors for the standard conjugation
  ($\eta = 1$)  and nonstandard pseudoconjugations  ($\eta=-1$).



\subsection{$N=2$ pseudoconjugations and corresponding 
 selfconjugate $N=2$ Euclidean superalgebras }  

We shall consider now the $N=2$ complex Euclidean superalgebra with the supercharges  $Q_{\alpha}^{a}$ and $\bar{Q}_{\dot{\alpha}a}$ (see Sect. 3.1). The pseudoconjugation map (\ref{3s312})
  for $N=1$ superalgebra can be extended to  $N=2$ as follows (see also \cite{IvanovZu,Castro,Buch}):
\begin{equation}\label{3s314}
(Q^{a}_\alpha)^{\ddagger}\;\;=\;\;\epsilon_{\alpha\beta}Q^{a}_\beta,\qquad
(\bar{Q}_{\dot{\alpha}a})^{\ddagger}\;\;=\;\;\eta\epsilon_{\dot{\alpha}\dot{\beta}}\bar{Q}_{\dot{\beta}a}\ ,
\qquad \eta=\pm 1 \, .
\end{equation}
In order to obtain the antiautomorphism of $N=2$ complex superalgebra (see Sec. 3.1) we should postulate
the relation  (\ref{x2.36}) 
   implying that  $q=0$ (see (A.1)) for $\eta=-1$ and $q=1$ for $\eta=1$.
  We get therefore two 
  prolongations to fermionic sector of the Euclidean conjugation (\ref{2a4}). 
The invariance of the relation (\ref{2s2101}--\ref{2s2111}) 
  under  the pseudoconjugation maps 
(\ref{3s314}) 
  implies the following reality conditions
for central charges $Z,\tilde{Z}$
\begin{equation}\label{3s317}
Z^{\ddagger}\;\;=\;\;(-1)^{q+1}Z,\qquad \tilde{Z}^{\ddagger}=(-1)^{q+1}\tilde{Z}
\end{equation}
i.e. we obtain two real ($q=1$) or imaginary ($q=0$) central charges.

It is easy to show that the   $N=2$  supersymmetrization of complex $\mathfrak{io}(4;C)$  algebra becomes selfconjugate under
pseudoconjugation (\ref{3s314}) 
 under the assumption that the  Euclidean fourmomentum generators ${\mathcal P}_\mu$ are 
real (see (\ref{2a5})). 
  Further, using the relations $\epsilon\bar{\sigma}_{i}\epsilon=\sigma_{i}$,
we should choose for $\mathfrak{o}(4,C)$ generators the Euclidean
  reality conditions.
The invariance of superalgebra under the pseudoconjugation (\ref{3s314}) 
  requires that (see (\ref{sp4}))
\begin{equation}\label{3s316}
{T_{i}}^{j}\;\;=\;\;({T_{i}}^{j})^{\dagger}\qquad {({t_{i}}^{j})^{d}}_{c}\;\;=-\;{({{\bar{t}\, _{i}}} ^{j})^{d}}_{c}\, ,
\end{equation}
where $t\mapsto\bar{t}$ is the complex conjugation of $2\times 2$ matrix elements describing fundamental realizations of $\mathfrak{gl}(2;\mathbb{C})$ generators $T^j_i$. If central charges  are absent the reality constraints (\ref{3s316}) 
  restrict the complex internal $\mathfrak{gl}(2;\mathbb{C})$ algebra  to its real form
$\mathfrak{gl}(2;\mathbb{R})$.   The presence of scalar central charges $Z, \tilde Z$
commuting with Euclidean $\mathfrak{o}(4)$ and internal symmetry generators reduces the internal symmetries $GL(2;R)$
to its subgroup $SL(2;R)$; i.e. the 
$\mathfrak{sl}(2;\mathbb{R})$ subalgebra of $\mathfrak{gl}(2;\mathbb{R})$ remains not broken for any value of central charges (\ref{3s317}) 
  and describes N=2 Euclidean  $R$-symmetry.

\subsection{$N=2$ conjugation and  corresponding $N=2$ real Euclidean superalgebra}

If $N=2$ it is possible to introduce as well the following  conjugation 
\begin{equation}\label{3s320}
(Q^{a}_{\alpha})^{\dagger}\;\;=\;\;\epsilon_{\alpha\beta} \epsilon^{ab} Q^b_{\beta },
\qquad (\bar{Q}_{{\dot\alpha}a})^{\dagger}\;\;=\;\;\eta\,\epsilon_{\dot\alpha\dot\beta}\epsilon_{ab} \bar{Q}_{\beta {b}}
\ ,\qquad \eta=\pm 1
\end{equation}
with involutive property
\begin{equation}\label{3s321}
((Q^a_\alpha)^\dagger)^{\dagger}\;\;=\;\;Q^{a}_{\alpha},\qquad((\bar{Q}_{\dot{\alpha}a})^{\dagger})^{\dagger}\;\;=\;\;
\bar{Q}_{\dot{\alpha}a} \, ,
\end{equation}
which is an antilinear antiautomorphism describing the superextension of conjugation (\ref{2a5}). 
  It can be checked that (\ref{3s320}) 
   describes conjugation if $\eta = 1$ and pseudoconjugation if  $\eta= -1$.
 Due to the relation  if $\eta=(-1)^q$ if $\eta=1$  (\ref{3s320}) 
 provides
   standard antiautomorphism of complex $N=2$ superalgebra 
	with $q=0$; 
	  if $\eta=-1$ one should postulate nonstandard antiautomorphism 
		 (see (A.1)) 
		 with $q=1$.
  The restrictions on $N=2$ central charges $Z, \tilde Z$ which follow from the isomorphism of $N=2$ Euclidean
 superalgebra under the (pseudo)conjugation (\ref{3s320}) 
  are the following
\begin{equation}\label{3x}
Z^\dagger= \eta\,Z\qquad \tilde Z^\dagger=\eta\,\tilde Z  , \qquad   \eta =\pm 1 \, .
\end{equation}
The covariance of relations  (\ref{x1a}) 
  under the conjugation (\ref{3s320}) 
	 implies that  the generators $T_A$ of internal symmetry $\mathfrak{gl}(2,C)$  satisfy the reality conditions  ($A=r,0$)
\begin{equation}\label{2x}
(T_r)^\dagger=T_r \qquad\quad T_0^\dagger =-T_0 \, . 
\end{equation}

One can express in $N=2$ case the relations (\ref{x1a}) 
   consistently with (\ref{3s320}) 
	 as follows
\begin{equation}\label{x}
    [T_0, Q^a_\alpha] = - {1\over 2}Q^a_\alpha \quad [T_r, Q^a_\alpha] = -{1\over 2}(\sigma_r)^a_{\ b} Q^b_\alpha \, .
\end{equation}
The generators $T_r$ and  $T_0$ due to relations (\ref{2x}) 
   and (\ref{x1}) 
  describe the internal $N=2$ algebra $\mathfrak{su}(2)\oplus \mathfrak{u}(1)=\mathfrak{u}(2)$ (see also e.g. \cite{Buch}).

In the realizations of superalgebra  permitting to define the  Hermitean conjugation of 
supercharges   
  $Q^a_\alpha \to Q^\star_{\dot{\alpha}a}$, \,  $\bar{Q}_{\dot{\alpha}a} \to \bar{Q}^{a \star}_{{\alpha}}$ the
antilinearity of  the automorphism  (\ref{3s320}) 
  can be realized explicitly. If  $\eta = 1$  (i.e.  $q=0$) one can introduce the following counterpart of the 
 conjugation (\ref{3s320})  which  
employs    the  Hermitean conjugation 
\begin{equation}\label{y3.25}
(Q^a_\alpha)^\dagger = \varepsilon_{\alpha\beta} \varepsilon^{ab} Q^{ b \star}_{\beta}
\qquad
(Q_{\dot{\alpha}a})^\dagger = \varepsilon_{\dot{\alpha}\dot{\beta}} 
\varepsilon_{ab} Q^{ \star}_{\dot{\beta}b} \, .
\end{equation}
In such a case  one can define  the following two  complex--conjugated  N=2 Euclidean superalgebras \cite{Luk}:

-- holomorphic N=2 Euclidean superalgebra 
${\mathcal{E}}(4; 2|\mathbb{C})$ generated by supercharges $Q^a_\alpha, \bar{Q}_{\dot{\alpha}a}$

-- antiholomorphic N=2 Euclidean superalgebra
 $\overline{{\mathcal{E}}(4; 2|\mathbb{C})}$
 generated by Hermitean -- conjugated supercharges 
$(Q^a_\alpha)^\star, (\bar{Q}_{\dot{\alpha}a})^\star$.

The reality conditions described by  conjugation 
(\ref{y3.25}) 
  maps
	 ${\mathcal{E}}(4; 2|\mathbb{C}) \longrightarrow \overline{{\mathcal{E}}(4; 2|\mathbb{C})}$, i.e.
	they describe  the inner automorphism of real N=2 Euclidean superalgebra
	 ${\mathcal{E}}(4; 2|\mathbb{C})\oplus \overline{{\mathcal{E}}(4; 2|\mathbb{C})}$.
	Such form of N=2 Euclidean reality conditions has been employed  
	 in earlier applications, e.g.  for the
	description of N=2 Euclidean supersymmetric field--theoretic models,
	formulated in  complex N=2  superspace (see e.g. \cite{ZuminoB, Luk,ButInvLod})


\subsection{$N=2$ real Kleinian   superalgebra with central charges}

For $N=2$ Kleinian real supersymmetry we have  the following possible reality conditions:
\begin{equation}\label{k2s1}
 ( Q^{a}_\alpha)^\oplus = Q^{a}_\alpha ,
  \qquad
  (\bar Q_{{\dot{\alpha}} a})^\oplus =  \eta\bar Q_{{\dot{\alpha}} a},\, \quad \eta=\pm 1
  \end{equation}
 which extend  from  $N=1$  to N=2 Kleinian reality condition  (\ref{sp2_0k}). 
 The $N=2$ Kleinian superalgebra for $q=0$ because of  the condition  $\eta=(-1)^q$ takes the following form 
 (see also (\ref {kk2}--\ref{ykk2}))

  \begin{eqnarray}{}\label{k2s2}
  \left\{ Q^a_\alpha , \bar Q_{\dot{\beta\,b}}\right\}
 & = &
 2\, \delta^a_b\, (\tilde\sigma_\mu)_{\alpha \dot{\beta}} \tilde{{\mathcal P}}^\mu
  \\[-6pt] \nonumber
  \\[-6pt] \label{k2s2a}
   \left\{  Q^a_\alpha , Q^b_{{\beta}}\right\}=\epsilon_{\alpha\beta}\epsilon^{ab}Z\, ,
  &  &   \left\{ \bar Q_{\dot{\alpha}a} , \bar Q_{\dot{\beta}b}\right\} = \epsilon_{\dot{\alpha}\dot{\beta}}\epsilon_{ab}\tilde Z \, ,
   \end{eqnarray}
where  the reality conditions for two central charges look as follows
\begin{equation}\label{y3x}
Z^\oplus= \eta\,Z\qquad \tilde Z^\oplus=\eta\,\tilde Z \, .
\end{equation}
The internal symmetry generator $T_A$  $(A=0, r)$  (see (\ref{x1}), (\ref{xx})) 
 satisfy   for vanishing central charges  the reality condition
 $$(T_A)^\oplus=T_A\, .$$
If central charges do not vanish the R-symmetry  $GL(2;\mathbb{R})$
 is reduced to  $SL(2;\mathbb{R})$. 

\section{Classical real Poincar\'{e} and Euclidean r-matrices and their N=1 superextensions}
\sec
\subsection{General remarks}

We shall follow the method used in our previous paper \cite{BLMT} based on  the following steps:

i) Consider Zakrzewski list of 21 real classical $r-$matrices satisfying classical Yang-Baxter (YB)
\footnote{From the list of Zakrzewski's 21 cases of Poincar\'{e} $r-$matrices, given 
 in \cite{Z94} only one set, denoted with
$\mathcal{N}=6$, does not satisfy homogeneous classical YB equation, i.e. cannot be lifted to twisted Hopf algebra.}, and use in their
presentation   the canonical Poincar\'{e} basis (Sec. 2.1);

ii) Remove the Poincar\'{e} reality conditions (see (\ref{2a3}) or (\ref{bor4.1})) 
  imposed in \cite{Z94}.
The generators $h, e_\pm, h', e'_\pm, P_1, P_2, P_\pm$
are becoming complex and we obtain corresponding  class of classical $r-$matrices for complex inhomogeneous $\mathfrak{io}(4,C)$ algebra. 

iii) Extend supersymmetrically the complex classical $r-$matrices obtained in ii) 
 to N=1 and  N=2 by adding suitable terms 
which depend on  supercharges $Q^a_\alpha, \bar Q_{\dot\alpha a}$  ($a=1\ldots N$). For N=2 we consider as well terms in classical r-matrix which depend on complex $N=2$ central charges $Z, \tilde Z$
   (see (\ref{2s2111}))
  in such a way that the supersymmetric N=2 complex $r-$matrices  satisfy  the classical super-YB equation.


In Sect.~4.2 we recall the Zakrzewski list of D=4 real Poincar\'{e} 
 and D=4 self--conjugate Euclidean
 r-matrices from
\cite{Z94}  as well as  D=4 \ N=1 real super--Poincar\'{e} 
   and pseudoreal super-Euclidean  r-matrices obtained in \cite{BLMT}. 
In Sect.~5 we present new results for supersymmetric r-matrices with standard N=2 Poincar\'{e} reality conditions (see Sect.~3.2) and N=2 Euclidean (pseudo)reality conditons (see Sect.~3.3--4). 

\subsection{D=4 Poincar\'{e} real r-matrices}

Let us  present the real D=4 Poincar\'{e} r-matrices listed in \cite{Z94} (see also \cite{T07}).
Using the decomposition of  $r \in \mathfrak{io}(3,1)\wedge \mathfrak{io}(3,1)$
\begin{equation}\label{luk1}
r= a  + b +c
\end{equation}
where ($\mathbb{P}$ denotes the fourmomenta generators)
\begin{equation}\label{luk2}
a \in \mathbb{P} \wedge \mathbb{P} \qquad
 b \in \mathbb{P} \wedge \mathfrak{o}(3,1) 
\qquad 
c \in \mathfrak{o}(3,1) \wedge \mathfrak{o}(3,1)\, .
\end{equation}
Zakrzewski \cite{Z94}  obtained the following list 

{\scriptsize
\begin{table}[h]
{\scriptsize 
\begin{center}
\begin{tabular}{ccccc}
\hline $c$ & $b$ & $a$ & $\#$ & ${\cal N}$\\ 
\hline $\gamma h'\wedge h$ & $0$ & $\alpha P_{+}\wedge P_{-}+\tilde{\alpha}P_{1}\wedge
P_{2}$ & $2$ & $1$\\
\hline $\gamma e'_{+}\wedge e_{+}$ & $\beta_{1}b_{P_{+}}^{}+\beta_{2}P_{+}\wedge h'$ &
$0$ & $1$ & $2$\\
$$ & $\beta_{1} b_{P_{+}}^{}$ & $\alpha P_{+}\wedge P_{1}$ & $1$ & $3$\\
$$ & $\gamma\beta_{1}(P_{1}\wedge e_{+}+P_{2}\wedge e'_{+})$ & $P_{+}\wedge(\alpha_{1}
P_{1}\!+\alpha_{2}P_{2})-\gamma\beta_{1}^2P_{1}\wedge P_{2}$ & $2$ & $4$\\
\hline $\gamma(h\wedge e_{+}$ & $$ & $$ & $$ & $$\\
$-h'\wedge e'_{+})$ & $0$ & $0$ & $1$ & $5$\\
$+\gamma_{1}e'_{+}\wedge e_{+}$ & $$ & $$ & $$ & $$\\
\hline $\gamma h\wedge e_{+}$ & $\beta_{1}b_{P_{2}}^{}+\beta_{2}P_{2}\wedge e_{+}$ & $0$
& $1$ & $6$\\
\hline $0$ & $\beta_{1}b_{P_{+}}^{}+\beta_{2}P_{+}\wedge h'$ & $0$ & $1$ & $7$\\
$$ & $\beta_{1}b_{P_{+}}^{}+\beta_{2}P_{+}\wedge e_{+}$ & $0$ & $1$ & $8$\\
$$ & $P_{1}\wedge(\beta_{1}e_{+}+\beta_{2}e'_{+})\,+$ & $\alpha P_{+}\wedge P_{2}$ &
$2$ & $9$\\
$$ & $\beta_{1}P_{+}\wedge(h+\sigma e_{+}),\;\sigma=0,\pm1$ & $$ & $$ & $$\\
$$ & $\beta_{1}(P_{1}\wedge e'_{+}+P_{+}\wedge e_{+})$ & $\alpha_{1}P_{-}\wedge
P_{1}+\alpha_{2} P_{+}\wedge P_{2}$ & $2$ & $10$\\
$$ & $\beta_{1} P_{2}\wedge e_{+}$ & $\alpha_{1}P_{+}\wedge P_{1}+\alpha_{2}
P_{-}\wedge P_{2}$ & $1$ & $11$\\
$$ & $\beta_{1} P_{+}\wedge e_{+}$ & $P_{-}\!\wedge(\alpha P_{+}\!+\!\alpha_{1}P_{1}\!+\!
\alpha_{2}P_{2})\!+
\tilde{\alpha} P_{+}\!\wedge P_{2}$ & $3$ & $12$\\
$$ & $\beta_{1} P_{0}\wedge h'$&$\alpha_{1}P_{0}\wedge P_{3}+\alpha_{2}P_{1}\wedge P_{2}$&
$2$ & $13$\\
$$ & $\beta_{1} P_{3}\wedge h'$&$\alpha_{1}P_{0}\wedge P_{3}+\alpha_{2}P_{1}\wedge P_{2}$&
$2$ & $14$\\
$$ & $\beta_{1} P_{+}\wedge h'$&$\alpha_{1}P_{0}\wedge P_{3}+\alpha_{2}P_{1}\wedge P_{2}$&
$1$ & $15$\\
$$ & $\beta_{1} P_{1}\wedge h$&$\alpha_{1}P_{0}\wedge P_{3}+\alpha_{2}P_{1}\wedge P_{2}$ &
$2$ & $16$\\
$$ & $\beta_{1} P_{+}\wedge h$&$\alpha P_{1}\wedge P_{2}+\alpha_{1}P_{+}\wedge P_{1}$ &
$1$ & $17$\\
$$ & $P_{+}\wedge(\beta_{1} h+\beta_{2} h')$&$\alpha_{1} P_{1}\wedge P_{2}$ & $1$ & $18$\\
\cline{2-5}
$$ & $0$ & $\alpha_{1} P_{1}\wedge P_{+}$ & $0$ & $19$\\
$$ & $$ & $\alpha_{1} P_{1}\wedge P_{2}$ & $0$ & $20$\\
$$ & $$ & $\alpha_{1} P_{0}\wedge P_{3}+\alpha_{2}P_{1}\wedge P_{2}$ & $1$ & $21$\\
\hline
\end{tabular}\end{center}
}
\caption{
Real D=4  (pseudo)real Poincar\'{e} r-matrices (all satisfy homogeneous CYBE  except $\mathcal{N}$=6).}
\end{table}}
where we use Cartan--Chevaley basis for $\mathfrak{o}(3,1)$ (see (\ref{bor4.1}--\ref{bor4.2})) and
 $P_{\pm}=P_{0}\pm P_{3}$. 
 Besides  $b_{P_{+}}^{}$, $b_{P_{2}}^{}$ are given by the
expressions:
\begin{eqnarray}\label{cr3}
\begin{array}{rcl}
b_{P_{+}}^{}\!\!&=\!\!&P_{1}\wedge e_{+}-P_{2}\wedge e'_{+}+P_{+}\wedge h~,
\\[6pt]
b_{P_{2}}^{}\!\!&=\!\!&2P_{1}\wedge h'+P_{-}\wedge e'_{+}-P_{+}\wedge e'_{-}~,
\end{array} %
\end{eqnarray}
and  provide r-matrices describing light-cone ($b_{P_+}$) and tachyonic ($b_{P_2}$) $\kappa$-deformation \cite{BallHerr, KosinskiMaslanka}.

\subsection{N=1 D=4 (pseudo) real  Poincar\'{e} and Euclidean supersymmetric r-matrices}

In \cite{BLMT} we have presented the list of possible supersymmetric D=4 N=1 super--Poincar\'{e} r-matrices $r^{(1)}$. 
 It appears that only  7 out of 21 classes of r-matrices present in Table~1 can be supersymmetrized.
 The N=1 super--Poincar\'{e} r-matrices  
 decomposed as follows:
\begin{equation}\label{luk4.4}
r^{(1)} = r + s=a+b+c+s  \, ,
\end{equation}
where $s\in \mathbb{Q}^{(1)} \wedge \mathbb{Q}^{(1)}$  ($\mathbb{Q}^{(1)}$ denote N=1 Poincar\'{e}
 supercharges).  The list of  r-matrices (\ref{luk4.4}) looks as follows:

{\scriptsize
\begin{table}[h]
{\scriptsize
\begin{center}
\begin{tabular}{cccccc}
\hline\smallskip $c$ & $b$ & $a$ & $s$ & ${\cal N}$
											\\
\hline $$$$$$$$$$ \\
 $\gamma e'_{+}\wedge e_{+}$ & $\beta_{1}b_{P_{+}}^{}+\beta_{2}P_{+}\wedge h'$ &
$0$ & $\beta_{1}\bar{Q}_{\dot 1}\wedge Q_{1}$ & $2$\smallskip\\
$ $ & $\beta_{1} b_{P_{+}}^{}$ & $\alpha P_{+}\wedge P_{1}$ &
$\beta_{1}\bar{Q}_{\dot 1}\!\wedge \!Q_{1} $
 & $3$\\
\hline$$$$$$$$$$\\
\par$\gamma h\wedge e_{+}$ & $\beta_{1}b_{P_{2}}^{}+\beta_{2}P_{2}\wedge e_{+}$ & $0$
& $i\beta_{1}(Q_{1}+\bar{Q}_{\dot 1})\wedge(Q_{2}-\bar{Q}_{\dot 2})$ & $6$\par\\

\hline$$$$$$$$$$\\
$0$ & $\beta_{1}b_{P_{+}}^{}+\beta_{2}P_{+}\wedge h'$ & $0$ &
$\beta_{1}\bar{Q}_{\dot 1}\wedge Q_{1}$ & $7$\\
$$ & $\beta_{1}b_{P_{+}}^{}+\beta_{2}P_{+}\wedge e_{+}$ & $0$ &
$\beta_{1}\bar{Q}_{\dot 1}\!\wedge\!Q_{1}$
& $8$\\
$$ & $P_{1}\wedge(\beta_{1}e_{+}+\beta_{2}e'_{+})+$ & $\alpha P_{+}\wedge P_{2}$ &
$\beta_{1}\bar{Q}_{\dot 1}\!\wedge\!Q_{1}$
& $9$\\
$$ & $\beta_{1} P_{+}\wedge h$ & $\alpha_2 P_{1}\wedge P_{2}+\alpha_{1}P_{+}\wedge P_{1}$ &
$\beta_{1}\bar{Q}_{\dot 1}\!\wedge\! Q_{1}$\!
& $17$\\
\hline
\end{tabular}\end{center}}
\caption{Real  D= 4  N=1 super--Poincar\'{e} r-matrices (all satisfy homogeneous  CYBE except $\mathcal{N}$=6).}
\end{table}}

We see that the superextension is realized in all cases except $\mathcal{N}$=6 with the help of unique supersymmetric term 
 $S= \beta_1 \bar{Q}_{\dot{1}} \wedge Q_1$, where  $\beta_1$ is purely imaginary,  
  which is invariant under the N=1 super--Poincar\'{e} conjugation.   
	
	The list of N=1 complex Euclidean supersymmetric r-matrices, which are  self-conjugate  under the pseudoconjugation 
	  (\ref{3s312}) 
	  with  $\eta= -1$ looks as follows: 


{\scriptsize
\begin{table}[h]
{\scriptsize
\begin{center}
\begin{tabular}{cccccc}
\hline\smallskip $c$ & $b$ & $a$ & $s$ & ${\cal N}$  \par \\
\hline 
&&&&&
\\
$\gamma h'\wedge h$ & $0$ & $\alpha P_{+}\wedge P_{-}\!+\tilde{\alpha}P_{1}
\wedge P_{2}$ & $\beta Q_{2}\wedge Q_{1}$ & $1$\smallskip\\
\hline 
\\
$$ & $\beta_{1}P_{0}\wedge h'$ & $\alpha_{1}P_{0}\wedge P_{3}+\alpha_{2}P_{1}\wedge P_{2}$&
 $\beta Q_{2}\wedge Q_{1}$ &$13$\\
$$ & $\beta_{1}P_{3}\wedge h'$ & $\alpha_{1}P_{0}\wedge P_{3}+\alpha_{2}P_{1}\wedge P_{2}$&
$\beta Q_{2}\wedge Q_{1}$ & $14$\\
$$ & $\beta_{1}P_{+}\wedge h'$ & $\alpha_{1}P_{0}\wedge P_{3}+\alpha_{2}P_{1}\wedge P_{2}$ &
$\beta Q_{2}\wedge Q_{1}$ & $15$\\
$$ & $\beta_{1} P_{1}\wedge h$ & $\alpha_{1}P_{0}\wedge P_{3}+\alpha_{2}P_{1}\wedge P_{2}$ &
$\beta Q_{2}\!\wedge\! Q_{1}$ & $16$\\
\cline{2-6}$$$$$$$$\\
$$ & $0$ & $\alpha_{1}P_{1}\wedge P_{+}$ & $\eta^{\alpha\beta}_{}Q_{\alpha}\wedge
Q_{\eta}$ & $19$\\
$$ & $$ & $\alpha_{2}P_{1}\wedge P_{2}$ & $\eta^{\alpha\beta}_{}Q_{\alpha}\wedge
Q_{\beta}$ & $20$\\
$$ & $$ & $\alpha_{1} P_{0}\wedge P_{3}+\alpha_{2}P_{1}\wedge P_{2}$ &
$\eta^{\alpha\beta}_{}Q_{\alpha}\wedge Q_{\beta}$ & $21$\\
\hline
\end{tabular}\end{center}}
\caption{
{Pseudoreal (selfconjugate under pseudoreality condition)  D= 4 \, N=1 super--Euclidean r-matrices (all satisfy homogeneous CYBE).}}
\end{table}}
\noindent
where $s \in Q^{(1)}_c \wedge Q^{(1)}_c$  and $Q^{(1)}_c$  denote N=1 complex Euclidean supercharges 
($Q_\alpha, \bar{Q}_{\dot{\alpha}}$) (see  (\ref{2s210}--\ref{2s212})). 
 Due to inner automorphisms of N=1 Euclidean superalgebra the selfconjugate  term can be chosen  only as
 $\beta Q_1 \wedge \bar{Q}_{\dot{1}}$, with  parameter $\beta$ purely imaginary.

We observe that among supersymmetric r-matrices in Table~2 the r-matrices with $\mathcal{N}$=2,3 and 7,8 contain term $b_{P_+}$ characterizing light-cone $\kappa$-deformation \cite{BallHerr}, and the super r-matrix $\mathcal{N}$=6 contains term  $b_{P_2}$ describing tachyonic 
$\kappa$--deformation \cite{KosinskiMaslanka}. 
It should be added that the standard ``time--like'' $\kappa$-deformation characterized by the term 
 $b_{P_0}$, is not present in Tables~1--3.
Further we comment that in Euclidean case without supersymmetry (Table~1)  and with N=1 supersymmetry 
(Table~3), the r-matrices characterizing $\kappa$-deformations are not present.

\section{N=2 extensions with central charges of N=1 super--Poincar\'{e} and super--Euclidean r-matrices}
\sec 

The N=2 superextensions of D=4  Poincar\'{e} and Euclidean r-matrices can be decomposed in the
following way compare with  formula  (\ref{cr3})
\begin{equation}\label{lukb5.1}
r^{(2)} = \tilde{a} + \tilde{b} + \tilde{c} + \tilde{s}  \, , 
\end{equation}
where $\tilde{a}, \tilde{b}, \tilde{c}$ 
contains contributions from central charges $\mathbb{Z} = (Z_1, Z_2)$  \,$(\mathbb{Z} \wedge \mathbb{Z} \equiv Z_1 \wedge Z_2)$

\begin{eqnarray}\label{lukb5.2}
&&\tilde{a}  \in \mathbb{P} \wedge \mathbb{P} 
\qquad  \nonumber
\\
&&\tilde{b} \in \mathbb{P} \wedge ( \mathfrak{o}(3,1) \oplus \mathbb{Z}) \nonumber 
\\
&& \tilde{c} \in \mathfrak{o}(3,1)\oplus \mathbb{Z} \wedge (\mathfrak{o}(3,1)\oplus \mathbb{Z})
\end{eqnarray}
and $\tilde{s} \in \mathbb{Q}^{(2)} \wedge \mathbb{Q}^{(2)}$,
 where $\mathbb{Q}^{(2)}$ denote N=2 supercharges.

We shall list in Sect.~5.1 the r-matrices 
  (\ref{lukb5.2}) 
	 invariant under N=2 Poincar\'{e} reality condition
   (\ref{sp2})--(\ref{sp3})  and in Sect.~5.2  the ones invariant  under the  pseudoconjugation
	 (\ref{3s314}) and conjugation (\ref{y3.25}).

\subsection{N=2 real super--Poincar\'{e} r-matrices}

Due to the reality condition  (\ref{sp3})
  N=2 Poincar\'{e} superalgebra depends on one complex supercharge
 $\mathbb{Z} \equiv Z_1 = Z^\dagger_2$.  
In such a case to every Zakrzewski r-matrix (see Table~1) one can add unique term bilinear in central
charges (see (\ref{lukb5.2}))
\begin{equation}\label{lukb5.3}
Z\wedge Z^\dagger \in \mathbb{Z} \wedge \mathbb{Z} \in \tilde{c} \, .
\end{equation}

It follows from Table~2 that only Zakrzewski r-matrices with ${\mathcal{N}}=2,3,6,7,8,9,17$
 admit $N=1$ supersymmetrization,  realized by universal term $\bar{Q}_1 \wedge Q_1$.
It appears that for N=2\  D=4 super--Poincar\'{e} r-marices the term $\tilde{s}$ in formula
   (\ref{lukb5.1})
	 is also universal and described by ($\alpha$ real, $\chi$, $\chi' = 0,\, \pm 1$)
\begin{equation}\label{lukb5.4}
\tilde{s} = \alpha (\chi \, Q^1_1 \wedge \bar{Q}^1_{\dot{1}} + \chi' \, Q^2_1 \wedge \bar{Q}^2_{\dot{1}}) \, .
\end{equation}
 The term (\ref{lukb5.4}) provides N=2 supersymmetrization of  r-matrices\footnote{If $\chi$ or $\chi'=0$ then we obtain only N=1 supersymmetrization.} described in Table~2.

If we denote $\tilde{b} = b+\Delta b$ 
   \, ($\Delta b \in \mathbb{P} \wedge \mathbb{Z}$;  see (\ref{lukb5.2}))  such a term is possible only for supersymmetric r-matrices if $N\geqslant  2$.
 These additional terms linearly dependent on central charges,  can be added  only for $\mathcal{N}=2,6$ by the following universal expression ($p_+ = p_0 + p_3; \  \beta$ complex)
\begin{equation}\label{lukb5.5}
\Delta b = p_+ \wedge (\beta \, Z + \bar{\beta} \, \bar{Z}) \, .
\end{equation}
For $\mathcal{N}=3,7,8,9,17$ one can add consistently with CYBE the following complex term 
  belonging to $\mathfrak{o}(3,1) \wedge \mathbb{Z} \in \tilde{c}$\,   ($\beta_1, \beta_2$ -- complex)
\begin{equation}\label{lukb5.6}
\Delta c = \beta_1 \, e_+ \wedge Z + \beta_2 \, e'_+ \wedge \, \bar{Z}\, .
\end{equation}
Unfortunately, term (\ref{lukb5.6}) can not satisfy the  Poincar\'{e} reality conditions.



Listed above supersymmetric  $r$-matrices can be presented
as a sum of subordinated $r$-matrices which are of super-Abelian and super-Jordanian types. The subordination
enables us to construct a correct sequence of quantizations and to obtain the
corresponding twists describing the quantum deformations. These twists are in general
case the super-extensions of the twists obtained in \cite{T07}.                            

\subsection{N=2 (pseudo) real super--Euclidean r-matrices}

Contrary to the N=2 super--Poincar\'{e} case, when there is only one conjugation providing reality
condition (see Sect.~3.2\footnote{We considered only here the reality structure defined by standard
antilinear antiinvolution  ($q=0$ in relation (A.1)) what corresponds to the choice $\eta=1$ in (\ref{sp2}).}),
 in N=2 case we have two types of reality structure:
\begin{description}
\item[a)] Defined by pseudoconjuation (\ref{3s314}) 
  (we consider $q=0$ and  $\eta=-1$),  
	 which is the straightforward extension of pseudoreality  structure considered for N=1  
	  in (\ref{3s312}).              
\item[b)] Defined by the conjugation (\ref{y3.25}). 
   We further assume  that the Hermitean conjugation
$Q^a_\alpha \to (Q^a_\alpha)^\star, \, \bar{Q}^a_{\dot{\alpha}} \to (\bar{Q}^a_{\dot{\alpha}})^\star$ is well defined.\footnote{Such situation occurs in supersymmetric QFT, with supercharges realized as differential operators on superspace fields.
 In fact the conjugataion (\ref{y3.25}) can be defined for suitable class of realizations of supercharges $Q^a_\alpha, \bar{Q}^a_{\dot{\alpha}}$.} 
 Then one can formulate N=2 Euclidean superalgebra in a Hermitean form if  we impose the subsidiary condition which follows from  (\ref{y3.25})  \,  ($\alpha = 1,2; \ a,b = 1,2$)
\end{description}
\begin{equation}\label{lukbb5.9}
Q^a_\alpha = (Q^a_\alpha)^\dagger  \Rightarrow  
 Q^a_{\, \alpha} = \varepsilon_{\alpha\beta} \varepsilon^{ab}(\bar{Q}^b_{{\beta}})^\star
\qquad
\bar{Q}^a_{\dot{\alpha}} = (\bar{Q}^a_{\dot{\alpha}})^\dagger   \Rightarrow
 \bar{Q}_{\dot{\alpha} a} = 
 \varepsilon_{\dot{\alpha} \dot{\beta}} \varepsilon_{ab} {Q}_{\dot{\beta} b}\,  ,
\end{equation}
or more explicitly 
\begin{eqnarray}\label{lukbb5.10}
Q^1_\alpha = \varepsilon_{\alpha \beta}(\bar{Q}^2_{\dot{\beta}})^\star       \qquad \qquad
Q^2_\alpha = -   \varepsilon_{\alpha \beta}(\bar{Q}^1_{\dot{\beta}})^\star \, ,
\\ \nonumber
\bar{Q}^1_{\dot{\alpha}} =   \varepsilon_{\dot{\alpha} \dot{\beta}}(Q^2_\beta)^\star
   \qquad \qquad  
	 \bar{Q}^2_{\dot{\alpha}} = -  \varepsilon_{\dot{\alpha} \dot{\beta}}(Q^1_\beta)^\star \, . 
\end{eqnarray}
We see that 
 the supercharges
 $Q^2_\alpha,  \bar{Q}^2_{\dot{\alpha}}$  can be expressed  by 
 $(\bar{Q}^1_{\dot{\alpha}})^\star $, $(Q^1_\alpha)^\star$ and N=2 superalgebra  can be  described by two
 pairs of complex Hermitean--conjugated supercharges which equivalently can be expressed as  Hermitean--conjugated pair of  four--component Dirac spinors (A=1,2,3,4)

\begin{equation}\label{lukbb5.12}
\Psi_A = (Q^1_\alpha, \bar{Q}^1_{\dot{\alpha}}),
\qquad \qquad
\Psi^\star_A = ((Q^1_\alpha)^\star, (\bar{Q}^1_{\dot{\alpha}})^\star) \, .
\end{equation}

We shall consider below separately the N=2 Euclidean r-matrices selfconjugate under pseudoconjugation  (\ref{3s314}) and conjugation (\ref{lukbb5.9}).

\subsubsection{N=2   super--Euclidean r-matrices selfconjugate under pseudoconjugation}

We consider the complex N=2 $\mathcal{\epsilon}(4; 2|\mathbb{C})$ r-matrices which are invariant under the map (\ref{3s314}); 
  we choose the standard version of  formula (A.1) with $q=0$ what implies $\eta = -1$.
For N=2 we should take into consideration only the Poincar\'{e} r-matrices from Table~1 with $\mathcal{N}=1,13-16, 19-21$ (see also Table~3) which   allow N=1 supersymmetrization.
 If we consider the relations (\ref{3s317})  
 with $q=0$,  we get the pair of independent  N=2 Euclidean central charges $Z$, $\tilde{Z}$ which are purely imaginary.
 They provide Euclidean counterpart of formula (\ref{lukb5.3}) describing universal contribution to N=2 super-r-matrices. 

The fermionic part $\tilde{s}_E$ of  N=2 super--Euclidean r-matrix (\ref{lukb5.1}) 
   for $\mathcal{N}=1, 13-16$ is described by the following pair of  two 
	forms bilinear in supercharges ($\alpha_1, \alpha_2, \tilde{\alpha}_1, \tilde{\alpha}_2$ are real)

\begin{eqnarray}\label{lukbb5.13}
\tilde{s}_{E;1} = \alpha_1 \, Q^1_1 \wedge Q^1_2 + \alpha_2 \bar{Q}^2_1 \wedge \bar{Q}^2_2 \, ,
\\ \nonumber
{\tilde{s}}_{E;2} = \tilde{\alpha}_1 \, Q^2_1 \wedge Q^2_2 +  \tilde{\alpha}_2 \bar{Q}^1_{\dot{1}} 
\wedge \bar{Q}^1_{\dot{2}} \, .
\end{eqnarray}
To either of two terms (\ref{lukbb5.13}) 
  one can add the unique term 
$\Delta c_E \in \mathfrak{o}(3,1) \wedge \mathbb{Z}$ which takes the form ($\alpha$ complex, $\beta_1, \beta_2$ real)

\begin{equation}\label{lukbb5.14}
\Delta c_E = (\alpha h + \bar{\alpha} h' ) (\beta_1 \, Z + \beta_2 \, \tilde{Z})\, .
\end{equation}

For $\mathcal{N}=19-21$  we can again choose pair of the purely fermionic terms $\tilde{s}_E$, which are described by formulae (\ref{lukbb5.13}); 
  the terms $\Delta c_E$, linear in $Z, \tilde{Z}$, are however not universal, different for three cases $\mathcal{N}=19,20$ and 21.

\subsubsection{N=2 real super-Euclidean r-matrices}

In this case the algebraic structure does not have a counterpart in the formulae obtained by Euclidean N=1 supersymmetryzation (see Table~3).
 The task consists in finding such N=2 complex  $\mathcal{\epsilon}(4; 2|\mathbb{C})$
   r-matrices which are consistent with  N=2 super--Euclidean reality conditions 
(\ref{lukbb5.10}). 
  We should mention that  it is necessary to consider the N=2 supersymmetrization of  Poincar\'{e} r-matrices for  all 
	$\mathcal{N}=1 \ldots 21$. 
	We obtain the following list of fermionic and central charge dependent  terms for varius choices of $\mathcal{N}$
	\\
	\\
	\noindent
	{\textbf {$\alpha$)}} $\mathcal{N}=1, 13-16, 18$

There are possible the following  independent four fermionic two-forms $\tilde{s}_{E;k}$  \ ($k=1,2,3,4;$ $\alpha_1, \ldots  \alpha_6$ real)  

\begin{eqnarray}\label{lukbbb5.14}
&&\tilde{s}_{E;1} = \alpha_1(Q^1_1 \wedge Q^1_2 + \bar{Q}^2_1 \wedge \bar{Q}^2_2)\, ,
\\ \nonumber
&& \tilde{s}_{E;2} = \alpha_2(Q^2_1 \wedge Q^2_2 + \bar{Q}^1_{\dot{1}} \wedge  {Q}^1_{\dot{2}})\,,
\\ \nonumber
&&\tilde{s}_{E;3} = i \alpha_3 \, Q^1_1 \wedge  \bar{Q}^2_{\dot{2}} 
 + i \alpha_4 {Q}^1_{{2}} \wedge  \bar{Q}^2_{\dot{1}}\, ,
\\ \nonumber
&&\tilde{s}_{E;4} = i \alpha_5 \, Q^2_1 \wedge  \bar{Q}^1_{\dot{2}} 
 + i  \alpha_6 {Q}^2_{{2}} \wedge  \bar{Q}^1_{{1}} \, .
\end{eqnarray}

The additional  bosonic terms   $\Delta c_E \in \mathfrak{o}(3,1)\wedge \mathbb{Z}$  which are linear in real central charges $Z, \tilde{Z}$ satisfying  (for 
$\eta=1$)  the reality conditions (\ref{3x})  is described again by formula (\ref{lukbb5.14}). 
\\
\\
\noindent
{\textbf{$\beta$)}} $\mathcal{N}$=19--21.

For such values of $\mathcal{N}$ one can add any of four three--parameter term 
${\tilde{\tilde{s}}}_{E;k}$  below,  bilinear 
in supercharges ($\alpha_1, \ldots \alpha_{12}$ real):

\begin{equation}\label{lukbb5.16}
\begin{array}{l}
{\tilde{\tilde{s}}}_{E;1} =  
\alpha_1 (Q^1_1 \wedge Q^1_1 + \bar{Q}^2_{\dot{2}} \wedge \bar{Q}^2_{\dot{2}})
+ 
\alpha_2 (Q^1_1 \wedge Q^1_2 + \bar{Q}^2_{\dot{1}} \wedge {Q}^2_{\dot{1}})
+
\alpha_3 (Q^1_2 \wedge Q^1_2 + \bar{Q}^2_{\dot{1}} \wedge \bar{Q}^2_{\dot{1}}) \, ,
\\
\\
{\tilde{\tilde{s}}}_{E;2} =  
\alpha_4 (Q^1_2 \wedge Q^1_2 + {Q}^2_{\dot{1}} \wedge \bar{Q}^2_{\dot{1}})
+ 
\alpha_5 (Q^2_1 \wedge Q^2_2 + \bar{Q}^1_{\dot{1}} \wedge \bar{Q}^1_{\dot{2}})
+
\alpha_6 (Q^2_2 \wedge Q^2_2 + \bar{Q}^1_{\dot{1}} \wedge \bar{Q}^1_{\dot{1}}) \, ,
\\
\\
{\tilde{\tilde{s}}}_{E;3} =  
\alpha_7 \, Q^1_1 \wedge  \bar{Q}^2_{\dot{2}} 
+ 
\alpha_8\,  Q^1_2 \wedge  \bar{Q}^2_{\dot{1}} 
+
\alpha_9  (Q^1_1 \wedge \bar{Q}^2_{\dot{1}} + {Q}^1_{{2}} \wedge \bar{Q}^2_{\dot{2}}) \, ,
\\
\\
{\tilde{\tilde{s}}}_{E;4} =  
\alpha_{10} \, Q^2_1 \wedge  \bar{Q}^1_{\dot{2}} 
+ 
\alpha_{11}\,  Q^2_2 \wedge  \bar{Q}^1_{\dot{1}} 
+
\alpha_{12}  (Q^2_1 \wedge \bar{Q}^1_{\dot{1}} + {Q}^2_{{2}} \wedge \bar{Q}^1_{\dot{2}}) \, .
\end{array}
\end{equation}

The  additional bosonic terms $\Delta b_E$ linear in central charges and fourmomenta are different for considered three classes of r-matrices,  namely 
 ($\beta_1, \beta_2, \gamma_2, \gamma_3, \gamma_0$ real)

\begin{eqnarray}\label{lukbb5.17}
&N= 19 : \   &\Delta b_E = p_1 \wedge (\beta_1 \, Z  + \beta_2 \, \tilde{Z})\, ,
\\ \nonumber
&N=20: \  &\Delta b_E = (p_1 +  \gamma_2 p_2) \wedge (\beta_1 Z + \beta_2 \tilde{Z})\, ,
\\ \nonumber
&N=21: \  &\Delta b_E = (p_1 + \gamma_2 p_2 + \gamma_3 p_3 + i \gamma_0 p_0) \wedge
 (\beta_1 Z + \beta_2 \tilde{Z}) \, .
\end{eqnarray}

Finally we point out that among all 21 Zakrzewski classes of  complex $\mathfrak{io}(4;\mathbb{C})$  r-matrices  only for $\mathcal{N}$=5 we could not find 
 the superextension  
 of complex $\mathfrak{o}(4;\mathbb{C})$ r-matrices 
 to 
 $\mathcal{\epsilon}(4; N|\mathbb{C})$  ($N=1,2$)  supersymmetric  r-matrices, i.e. we were not able  for $\mathcal{N}$=5 to provide any consistent fermionic term 
 $\tilde{s}$ in  (\ref{lukb5.1}).

\section{Final remarks}

This paper provides firstly systematic discussion of real forms of  $\mathfrak{o}(4;\mathbb{C})$,
 $\mathfrak{io}(4; \mathbb{C})$,  $\mathcal{\epsilon}(4; 1|\mathbb{C})$ and
 $\mathcal{\epsilon}(4;2| \mathbb{C})$,  where $\mathcal{\epsilon}(4;N; \mathbb{C})$ 
 describes complex D=4 N--extended Euclidean superalgebra.
 In Sect.~2 and~3 we consider the reality and pseudoreality conditions ( reality constraints).  To the Poincar\'{e} and Euclidean 
 (pseudo)real forms we added also the (pseudo)real forms for the  Kleinian signature $g_{\mu\nu}=\hbox{diag} (1,-1,1,-1)$.
 In particular we also  considered  the (pseudo) reality conditions leading to exotic supersymmetry scheme, with odd (Grassmann) coordinates conjugated in nonstandard way (see e.g.
(\ref{x2.39})--(\ref{xy2.39}), with $q=1$). 

Our second aim was to present the extension of Zakrzewski list of classical D=4 Poincar\'{e} r-matrices to Euclidean case and N=1,2 supersymmetrizations.
 The D=4 N=1 Poincar\'{e} and Euclidean supersymmetric classical r-matrices already considered in 
\cite{BLMT} were presented in Sect.~4.
 In Sect.~5 we describe partial results for the classification of D=4 \ N=2 supersymmetric r-matrices for various D=4 signatures.
 We consider in this paper N=2  Poincar\'{e} and Euclidean  signatures  and these results are new.
 It should be pointed out that we considered also the terms depending on a pair of N=2 central charges 
(for $\mathcal{\epsilon}(4;2| \mathbb{C})$);  they are complex-conjugated for Poincar\'{e} and 
 real for Euclidean   cases.

We did not  consider the deformations of  N=2 Kleinian supersymmetry and corresponding classical r-matrices.
More systematic approach, with more complete  list of complex classical $\mathcal{\epsilon}(4;2| \mathbb{C})$ r-matrices and their various real forms will be considered in our subsequent  publication.

\section*{Acknowledgments}
The paper was presented at XXIIIth International Conference on Integrable Systems and Quantum Symmetries, Prague (June 2015). 
 The presenting author (JL) would like to thank prof. Cestimir Burdik for warm hospitality.
 The paper has been supported by the Polish Science Center  (NCN), project 
 2014/13/B/ST2/04043 (A.B. and J.L.), RFBR grant No 14-01-00474-a (V.N.T.).


\appendix
\section{Conjugations and pseudoconjugations of complex Lie superalgebras}
\sec

  In order to describe the supersymmetries of physical  systems one should consider real and pseudoreal forms of complex Lie superalgebra $L$, which are defined  in algebraic framework with help of the conjugations and pseudoconjugations.  The conjugations and pseudoconjugations are  usually defined as abstract antilinear antiautomorphisms $x \to x^\star$ of second and fourth  order preserving the $Z_2$ grading of superalgebra and satisfying the properties ($x,y\in U_L$; $|x|=0$ ($|x|=1$) describes the parity of even (odd) element $x$), where $U_L$
  denotes enveloping algebra of $L$:
 \begin{eqnarray}\label{3s31c}
 (xy)^\star &=& (-1)^{q|x||y|}\,y^\star \, x^\star\qquad q=0,1
 \label{3s31} \\
 (\alpha x + \beta y)^\star & =& \bar{\alpha} x^\star +  \bar{\beta}  y^\star \qquad \quad
  \alpha, \beta \in C
   \label{3s32}
 \end{eqnarray}
 where $q=0, 1$ defines two types of antilinear antiinvolution map in $U_L$ and $\alpha \to \bar{\alpha}$, $\beta \to \bar{\beta}$ describe complex conjugation in $C$. Further
 \begin{subeqnarray}
 (x^\star)^\star = x  &  \qquad \quad \hbox{(conjugation}) \, ,&
 \label{3s33a}
 \\
 (x^\star)^\star = - x & \qquad \quad  \hbox{(pseudoconjugation})\, . &
 \label{3s3b}
 \end{subeqnarray}
For  Lie superalgebras the property (\ref{3s3b}b) 
  occurs only in odd parity sector, i.e. for any $x\in L$ both relations (\ref{3s33a}a) 
	and (\ref{3s3b}b) 
	can be written together as
$$
(x^\star)^\star = (-1)^{p|x|}x \, ,
\eqno{(A.3c)}
$$
where $p=0$ (resp. $p=1$) denotes conjugations (resp. pseduconjugation) and we recall that $|x|$ describe the grading of superalgebra element $x$.
The conjugations (\ref{3s33a}a) 
  in matrix and Hilbert space realizations of superalgebra can be identified with the Hermitean conjugation, and pseudoconjugations in odd sector of the matrix superalgebras with $p=1$ were introduced as graded Hermitean conjugation \cite{Scheunert} (see also \cite{BerVNT}).
In the case of conjugations (A.3a)  the  Hermitean elements  $L_R,  \widetilde{L}_R$ of complex Lie algebra $L$ are defined as follows
\begin{equation}\label{3ss34}
L_R \ : \ x_R =  (x +x^\star) \qquad
\widetilde{L}_R :\ \ \  \widetilde{x}_R =  i(x - x^\star) \qquad   x \in L \, ,
\end{equation}
where $x^+_R = - x_R$ and ${\tilde{x}}{}^+_R = - {\tilde{x}}_R$.
In the case of conjugation the superalgebras $L_R, {\widetilde{L}}_R$ are 
 the subsuperalgebras  which can be defined as fixed points of the conjugation map, and  $L=L_R\oplus \widetilde{L}_R$ provides the formula describing the realification of $L$.
In the case of pseudoconjugations (\ref{3s3b}b) 
  the elements (\ref{3ss34}) 
	 of complex Lie algebra
satisfy the set of relations
\begin{equation}\label{3ss36}
x^\#_R = - (x - x^\#) = i \widetilde{x}_R
\qquad \quad
\widetilde{x}_R^\# = - i x_R
\end{equation}
with elements $x_R, \widetilde{x}_R$  satisfying involutive relations of fourth order
\begin{equation}\label{3ss38}
(x_R^\#){}^\# = - x_R \qquad \qquad (x^\#_L){}^\# = - x_L \, .
\end{equation}
We comment  that one uses for description of real supersymmetries in classical physics  an alternative  conjugations and pseudoconjugations, which are antilinear  automorphism  with the property
\begin{equation}\label{3ss39}
\overline{xy} = \bar{x}\, \bar{y}\ \ \ ;
\end{equation}
the remaining relations (\ref{3s32}), (\ref{3s33a}a--b)  are
unchanged. One can add that authomorphisms (\ref{3ss39})
  for  classical physical systems are usually represented by complex conjugation, and antiauthomorphisms adjusted to quantum systems are realized as (graded) Hermitean conjugation
	 in suitable Hilbert space framework.

\end{document}